\documentclass[onecolumn]{aastex631}

\begin{document}

\title{On The Lunar Origin of Near-Earth Asteroid 2024 PT5}

\correspondingauthor{Theodore Kareta}
\email{tkareta@lowell.edu}

\author[0000-0003-1008-7499]{Theodore Kareta}
\affiliation{Lowell Observatory, 
Flagstaff, AZ, USA}

\author[0000-0001-5875-1083]{Oscar Fuentes-Mu\~{n}oz}
\affiliation{Jet Propulsion Laboratory, California Institute of Technology, 4800 Oak Grove Dr, Pasadena, CA 91109, USA}

\author[0000-0001-6765-6336]{Nicholas Moskovitz}
\affiliation{Lowell Observatory, 
Flagstaff, AZ, USA}

\author[0000-0003-0774-884X]{Davide Farnocchia}
\affiliation{Jet Propulsion Laboratory, California Institute of Technology, 4800 Oak Grove Dr, Pasadena, CA 91109, USA}

\author[0000-0003-1383-1578]{Benjamin N.L. Sharkey}
\affiliation{Department of Astronomy, University of Maryland \\
4296 Stadium Dr. \\
PSC (Bldg 415) Rm 1113 \\
College Park, MD 20742-2421, USA \\
}

\begin{abstract}
The Near-Earth Asteroid (NEA) 2024 PT5 is on an Earth-like orbit which remained in Earth's immediate vicinity for several months at the end of 2024. PT5's orbit is challenging to populate with asteroids originating from the Main Belt and is more commonly associated with rocket bodies mistakenly identified as natural objects or with debris ejected from impacts on the Moon. We obtained visible and near-infrared reflectance spectra of PT5 with the Lowell Discovery Telescope and NASA Infrared Telescope Facility on 2024 August 16. The combined reflectance spectrum matches lunar samples but does not match any known asteroid types -- it is pyroxene-rich while asteroids of comparable spectral redness are olivine-rich. Moreover, the amount of solar radiation pressure observed on the PT5 trajectory is orders of magnitude lower than what would expected for an artificial object. We therefore conclude that 2024 PT5 is ejecta from an impact on the Moon, thus making PT5 the second NEA suggested to be sourced from the surface of the Moon. While one object might be an outlier, two suggest that there is an underlying population to be characterized. Long-term predictions of the position of 2024 PT5 are challenging due to the slow Earth encounters characteristic of objects in these orbits. A population of near-Earth objects which are sourced by the Moon would be important to characterize for understanding how impacts work on our nearest neighbor and for identifying the source regions of asteroids and meteorites from this under-studied population of objects on very Earth-like orbits. 
\end{abstract}


\section{Introduction} \label{sec:intro}
A primary goal in the study of Near-Earth Objects (NEOs), and specifically the Near-Earth Asteroids (NEAs), has been to relate their current orbits and properties back to their sources -- primarily, though not exclusively, in the Main Asteroid Belt. This connection has been done on a case-by-case basis through comparison with meteorites since the 1970s (see, e.g. \citealt{1970Sci...168.1445M, 1977GeCoA..41.1271C}) and on a systematic level with population-scale dynamical modeling since the early 2000s \citep{2002Icar..156..399B, 2018Icar..312..181G, 2024Icar..41716110N}. Meteoritic comparisons require physical data about the object (e.g., its reflectance spectrum or visible albedo) but are affected by biases in terrestrial meteorite collections and by challenges in assessing the affects that grain size has on reflectance spectra (see, e.g., \citealt{2023PSJ.....4...52B, 2023PSJ.....4..177C}). While the NEO Source Region Models mentioned above have proven an increasingly critical tool in predicting where the NEOs came from, they often only provide statistical assessments for individual objects-- multiple sources might be able to put objects onto a given orbit. Furthermore, one must assume that these models are based on a complete assessment of the NEO population -- if there were populations which existed but which were poorly surveyed (or perhaps hard to survey \textit{for}), the models might produce confident-and-incorrect results about certain kinds of orbits. Caveats aside, both of these approaches appear to work well to explain large-scale NEA population trends from well-established transportation pathways from the main asteroid belt. In any case, these approaches are getting more reliable as we find more asteroids and measure more of their properties. Unique or rare objects, however, may have their origins misinterpreted by assuming that no new pathways exist. When possible, an approach based on both dynamics and observations is thus most effective in diagnosing a NEA's origin and in turn estimating other properties which are challenging to measure remotely -- like density, bulk composition, or the age of its surface.

There are few objects of any kind with orbits \textit{extremely} similar to the Earth. A common criterion to assess this would be the velocity-at-infinity or asymptotic relative velocity ($v_\infty$) of the object, i.e., the relative speed of the object compared to the Earth at their closest approach calculated as if the Earth were massless. While there are many asteroids with $v_\infty > 4.0$ km/s, there are few with lower values. For objects with $v_\infty$ that approaches or is lower than the escape or orbital speeds of the Moon ($\approx2.4$ km/s and $\approx1.0$km/s, respectively), there are thought to be two primary sources of objects. The first are artificial ones, such as pieces of rockets which are re-discovered later by NEO surveys and mistakenly classified as natural. In most cases, these kinds of objects have high area-to-mass ratios and thus have their orbits perturbed by radiation pressure on relatively short (weeks-to-months) timescales. This kind of large solar radiation perturbation was what lead to the assessment that asteroid 2018 AV2 was likely artificial, that J002E3 may be be the upper stage of \textit{Apollo 12}'s \textit{Saturn V}\footnote{JPL Press release of 2002 Sept 20: \url{https://www.jpl.nasa.gov/news/first-confirmed-capture-into-earth-orbit-is-likely-apollo-rocket/}}, or that 2020 SO was a rocket body from \textit{Surveyor 2} \citep{2024PSJ.....5...96B}. The other source of low $v_\infty$ objects is a natural origin in the Earth-Moon system, such as macroscopic pieces of ejecta from impacts. While ejecta from impacts on the Earth producing NEOs is technically plausible, the Earth's higher escape speed would thus require a larger (and thus rarer) impact to launch material into a heliocentric orbit compared to the Moon and thus lunar origins for these natural low-$v_\infty$ objects is more likely. For any object in this kind of orbit, it should undergo repeated slow close encounters with the Earth and thus should have significant orbital evolution on relatively short timescales for NEOs. Even in the rare scenario that a near-Earth asteroid is dynamically evolved enough to end up in a low $v_\infty$ orbit, it should not stick around for long before getting scattered somewhere else (see, e.g., \citealt{2022AJ....163..211Q}).

\subsection{Kamo`oalewa and the Mars Trojans}
\citet{2021ComEE...2..231S} presented observational and dynamical arguments for a lunar origin of the Earth Quasi-Satellite (469219) Kamo`oalewa. While a visible spectrum alone was consistent with both red L-type asteroids and lunar materials, the addition of near-infrared photometry revealed the object to be significantly redder than other near-Earth asteroids. While interpreting Kamo`oalewa's reflectance spectrum as a traditional asteroid was challenging, a comparison with returned lunar samples showed good agreement in its reflective behavior at all wavelengths studied. Those authors argued that given the lack of asteroids in such Earth-like orbits, the qualities of Kamo`oalewa's particular orbit, and the way it reflected light, an origin as ejecta from the lunar surface was more plausible than an origin in the Asteroid Belt. Subsequently, dynamical simulations have added more evidence in support of this hypothesis. Kamo`oalewa's orbit is unusually stable -- alternating between co-orbital and horseshoe orbits for many hundreds of thousand of years -- ejecta from impacts on the trailing hemisphere on the Moon may be able to produce objects in just such dynamical states \citep{2023ComEE...4..372C}. \citet{2024NatAs...8..819J} suggested that the crater Giordano Bruno on the Far Side of the Moon might be where the Quasi-Satellite was sourced from on a basis of the age of the crater (1-10 million years old), its placement on the lunar surface, and cratering simulations.

The potential for interdisciplinary collaborations between the communities that study the Moon and the NEOs is clearly significant. If Kamo`oalewa or any other lunar NEO could be linked to a specific crater on the Moon, its formation age would directly link to a lunar surface feature, and the asteroid's materials would provide direct insight into the mechanics of crater excavation that are perhaps not available otherwise -- what kinds of materials are ejected from what depths at what shock states at which velocities. Understanding if NEO population models require additional sources of small, near-Earth impactors is critical to properly link the asteroid population to the recent cratering chronology of the moon.

Some of the Trojan Asteroids of Mars (the Eureka cluster, named after their largest member) have been proposed \citep{2017NatAs...1E.179P} to be sourced from a significant impact on that planet  based on their reflectance spectra. Specifically, that study proposed that (5621) Eureka and other members of its `orbital cluster' were excavated from the Martian mantle due to their olivine-rich achondritic spectra, similar to some Martian meteorites proposed to be from that planet's mantle -- and thus likely sourced from a significant impact deep into the past. Some questions about the Martian mantle therefore, like the recent impact history of the Moon, might be easier to investigate by looking at the small bodies nearby than by looking at the objects themselves.

For both Kamo`oalewa and the Eureka cluster, aspects of the impact histories of the terrestrial planets are hidden among the asteroid populations that are close by. While obtaining physical information about these objects on very similar orbits to the planets is challenging -- Eureka is less than two kilometers across and Kamo'oalewa is a hundred meters across at most -- it is only through these physical observations that their importance was fully realized. It is thus of interest to the Small Body and terrestrial planet communities to continue to survey these populations and to discern their origins.

\section{Ground-based Observations} \label{sec:obs}
2024 PT5 was discovered on 2024 August 7 by the ATLAS \citep{2018PASP..130f4505T} project's South Africa station at an approximate orange filter magnitude of $17$ \citep{2024MPEC....P..170T}. The low delta-V Earth-like orbit of the object was apparent immediately which made PT5 a natural target for characterization for our Mission Accessible Near-Earth Object Survey (MANOS, see, e.g., \citealt{2016AJ....152..163T, 2019AJ....158..196D}.) We first characterized the object photometrically with the Lowell Discovery Telescope (LDT) on August 14th. One aim of these initial observations was to assess whether the reflectance properties of PT5 were indicative of a natural or artificial origin, a distinction that can be made based on the presence of a broad 1-micron silicate absorption band. These observations were then followed by activating simultaneous target-of-opportunity (ToO) observing programs at the LDT and the NASA Infrared Telescope Facility on the 16th. A broad quantitative summary of our observations is available in Table \ref{tab:obs}.

\startlongtable
\begin{deluxetable*}{c|c|c|c|c|c}
\tablecaption{Summary of Observations.}
    \label{tab:obs}
    \centering
    \startdata
        Target/Instrument/Date & $m_V$ & Airmass & Usable Time on Target & Phase Angle [$^\circ$] & Comments\\
        \hline
         2024 PT5/LMI/2024-08-14 & 18.0 & 1.13-1.19 & 882 s (VR), 126 s (colors) & 1.4 & Target asteroid.\\
         2024 PT5/DeVeny/2024-08-16 & 18.5 & 1.04-1.10 & $8 \times 120$ s $ = 960$ & 1.3-1.4 & Target asteroid.\\
         SA 110-361/SpeX/2024-08-16 & 12.4 & 1.22 & $5 \times 15$ s $ = 75$ & n/a & Solar Analog.\\
         2024 PT5/SpeX/2024-08-16 & 18.5 & 1.03-1.00 & $29 \times 120$ s $ = 3480$ & 1.2 & Target asteroid.\\
         SA 110-361/SpeX/2024-08-16 & 12.4 & 1.03-1.00 & $8 \times 3 \times 10$ s $ = 240$ & n/a & Solar Analog.
    \enddata
\end{deluxetable*}

We first obtained multi-filter photometric observations of the asteroid with the LDT's Large Monolithic Imager (LMI, \citealt{2014SPIE.9147E..2NB}) on the night of 2024 August 14. Conditions were good, with no clouds and stable seeing of $1.4$ arcseconds thoughout the observing sequence. A series of 126 images (7-second exposures) with a broad (0.5-0.7 micron) VR filter were taken to constrain lightcurve properties, and a series of exposures in the SDSS griz filter set were obtained to constrain spectro-photometric properties. A total of 26 griz exposures (again 7-second exposures) were obtained with the following filter sequence: r-r-g-g-r-r-i-i-r-r-z-z-r-r-g-g-r-r-i-i-r-r-z-z-r-r. The interweaved r-band measurements were obtained to track any lightcurve variability during the $\sim12$ minutes required to collect the color data. Conditions during these imaging observations were stable with average seeing FWHM around 2$"$.

All LMI data were bias subtracted and flat field corrected with dome flats before processing with the \textit{PhotometryPipeline} \citep{2017A&C....18...47M}, an all-in-one calibration and extraction pipeline optimized for photometric observations of Solar System small bodies. A fixed photometric aperture of 2.3$"$ in diameter, optimized to maximize signal-to-noise, was used for all images. We calibrated the extracted magnitudes of 2024 PT5 against $\sim50-90$ PanSTARRS field stars in each frame and, to ensure reliability of the calibrated magnitudes, we only used stars which had Sun-like colors (i.e., $g-r$ and $r-i$ colors within 0.2 mag of the Sun). 

\begin{figure}[ht!]
\plotone{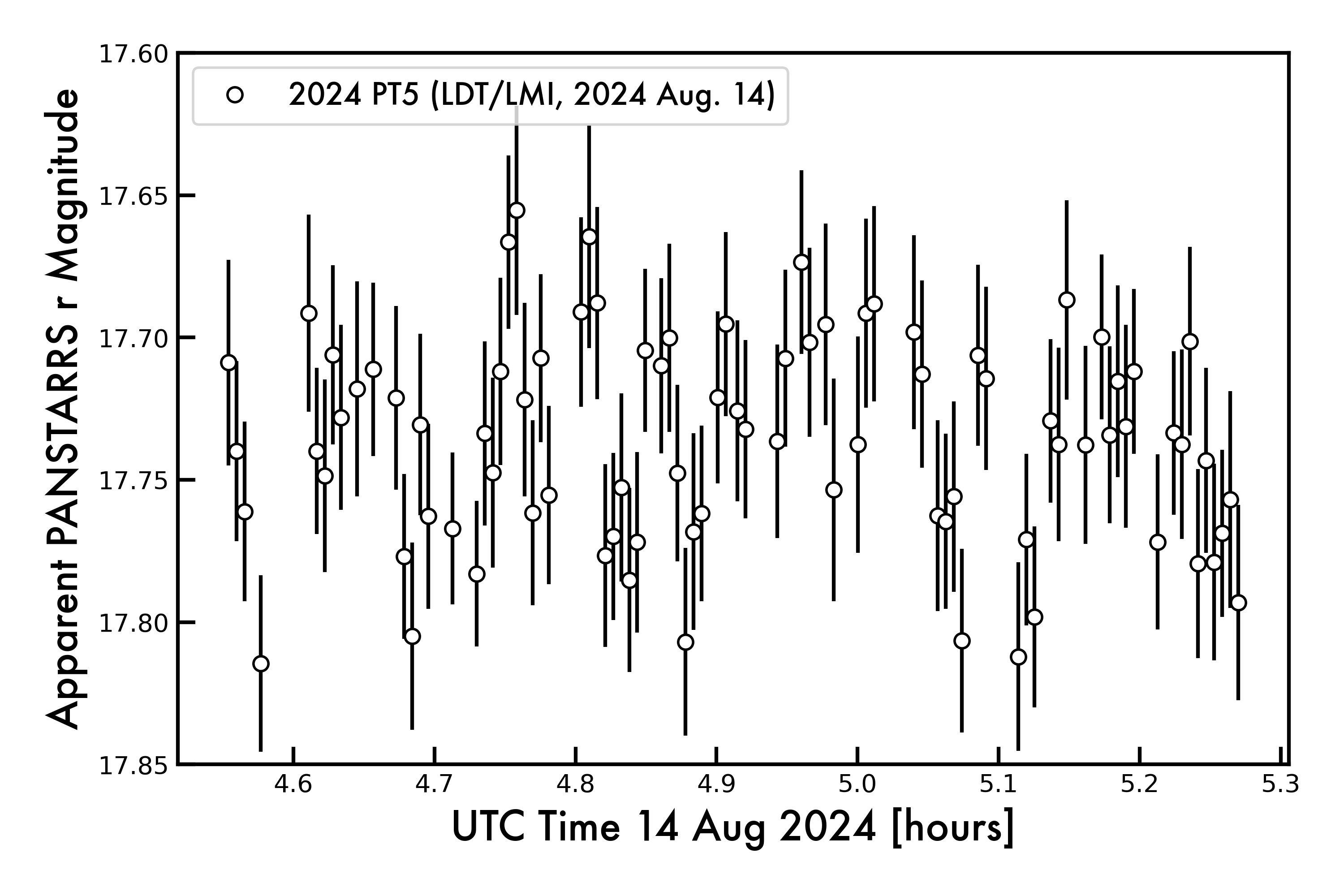}
\caption{Our 2024 August 14 photometric observations of 2024 PT5 are shown as black unfilled circles with one-sigma error bars. No obvious periodicity or variability is seen, suggesting that the lightcurve is either low-amplitude (implying a roughly circular shape of the object projected along the line of sight), very short ($<30$ seconds) or significantly longer than our lightcurve observations ($>40$ minutes).}
\label{fig:pt5_lc}
\end{figure}

Our lightcurve photometry was collected over a span of about 40 minutes and is shown in Figure \ref{fig:pt5_lc}. The VR images were calibrated to the PanSTARRS $r$-band and showed a mean and standard deviation of $r = 17.7~\pm~0.04$. Though some of the variability in this time series could be attributed to the rotational lightcurve, the data show no clear sign of a periodic signature. This suggests one or more possible interpretations including: slow rotation ($P>>40$ minutes), fast rotation ($P<<30$ seconds), non-principal axis rotation so that multiple frequencies may be present in the time series, or a viewing geometry such that the cross-sectional area of PT5 changed little (e.g. due to pole-on viewing or a spherical morphology) across the 40 minute observing window. Regardless of the object's rotational state the variations in the time series photometry are small and just above the formal error bars on the photometry. Thus no uniquely diagnostic information can be extracted about the rotational state of the body based on our observations.

\begin{figure}[ht!]
\plotone{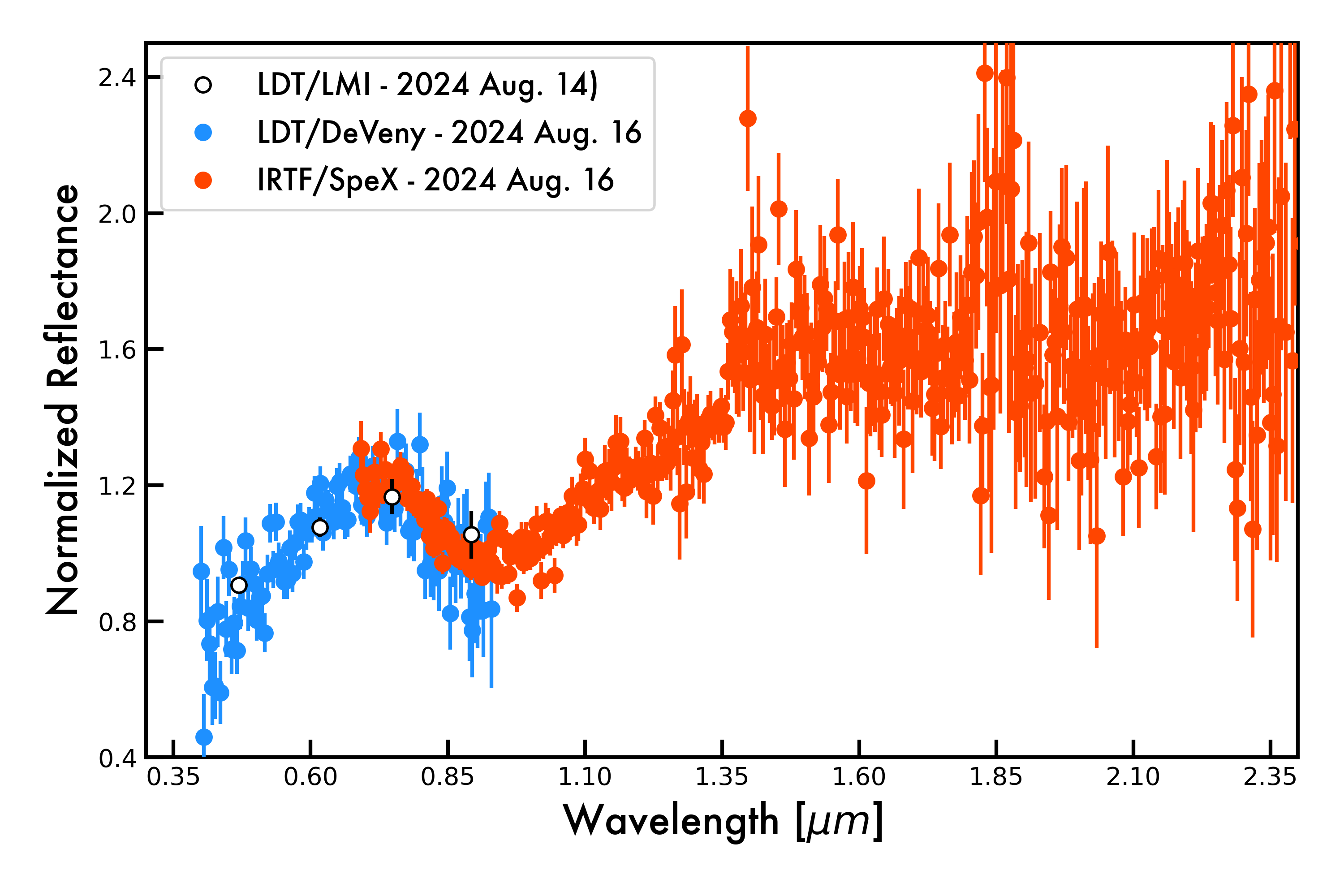}
\caption{Our visible photometric observations (taken with LMI on the LDT, white circles) of 2024 PT5 from August 14th are compared against the visible (Deveny on the LDT, blue circles) and near-infrared spectra (SpeX on the IRTF, red circles) taken on the 16th. The three datasets agree well in the region of overlap and high SNR ($0.75\mu{m} < \lambda < 0.90\mu{m}$). The surface of 2024 PT5 is likely silicate rich based on the presence of a strong $0.9\mu{m}$ absorption feature and a weak one at $2.0\mu{m}$.} 
\label{fig:reflectance}
\end{figure}

With no appreciable lightcurve variability, the determination of colors was relatively straightforward. We computed the mean r-band magnitude ($=17.75~\pm~0.03$) across the 12-minute color sequence and used that as a reference value to then compute colors relative to r-band for each exposure in the other filters. This resulted in average colors $g-r = 0.618\pm0.029$, $r-i = 0.205\pm0.043$, and $r-z = 0.064\pm0.083$, where the error bars represent the standard deviation of each set of individual color measurements. These average colors were converted to reflectivity by comparison with the colors of the Sun \citep{2018ApJS..236...47W}. The resulting spectro-photometry is shown as white dots in Figure \ref{fig:reflectance}. The indication of a 1-micron absorption band in the spectro-photometry suggests a natural origin for PT5 and thus motivated additional observations. We note that while our colors are similar with some slight differences than those recently reported by \citet{2024arXiv241108029B}, our lightcurve results are wholly different -- they saw significant variability over the timescale over which we saw little-to-none. While further observations of this object are clearly warranted to sort out this and other issues, it is possible that the object is in a tumbling state that made it have a more appreciable lightcurve during their observations compared to ours. PT5 was also significantly dimmer when their observations were taken compared to ours, which would add to the uncertainty in their assessment of the variability of the object's brightness. We return to a comparison of our observations with those obtained by others in Section \ref{sec:disc}.

Following those photometric observations, we activated Target-of-Opportunity programs to obtain visible spectroscopy of the target with the DeVeny instrument at the LDT and with SpeX \citep{2003PASP..115..362R} on the NASA IRTF simultaneously on 2024 August 16. The target was dimming steadily, so the time to obtain reflectance spectra with sufficient signal to noise was limited. These two instruments both produce spectra with comparable resolution ($R\sim$a few hundred, with DeVeny having finer resolution than SpeX) and thus are natural instruments for a paired visible/near-infrared study of a Solar System Small Body. At both sites, the conditions were also relatively good. At the LDT, seeing varied between $1.3$ and $1.4$ arcseconds over the observing sequence with varible light clouds. At the IRTF, the sky was clear and the seeing was slightly better at a stable $1.2$ arcseconds. We utilized a $0.8"$-wide slit on SpeX and a $3.0$"-wide slit on DeVeny and in both cases `book-ended' observations of the target with observations of a nearby Solar Analog star at very similar airmass. The SpeX observations were reduced within the \textit{spextool} pipeline \citep{2004PASP..116..362C} and the DeVeny observations were reduced within the \textit{Pypeit} environment \citep{2020JOSS....5.2308P}. The extracted spectra of the target were divided by the extracted spectra of the nearby Solar Analogs to produce our reflectance spectra. In the case of the visible wavelength DeVeny data, we also applied an extinction correction using an empirical extinction curve to each spectrum before combination and division, though this did little to reduce the scatter as the standard and target were observed at such similar airmasses. The LDT/DeVeny spectrum is shown in blue  and the IRTF/SpeX spectrum is shown in red in Figure \ref{fig:reflectance}. Due to the faintness of the target ($m_V \approx 18.5$), we binned both spectra to comparable SNR ($10\times$ binning for the visible spectrum, $0.005\mu{m}$ per pixel, $3\times$ binning for the near-infrared spectrum, $0.035\mu{m}$ per pixel). We combined the spectra into a single visible/near-infrared spectrum for further analyses described in the next Section at the wavelength resolution of the binned visible spectrum, but choice of binning approach or how to combine the datasets did not significantly impact any of our analyses.

As can be seen in Figure \ref{fig:reflectance}, the LMI-derived reflectance photometry taken on the 14th agrees well with the DeVeny derived reflectance spectrum taken on the 16th. Furthermore, the DeVeny and SpeX spectra agree well where their wavelengths overlap ($0.7-0.92\mu{m}$). This adds confidence that each of the three datasets was reduced and calibrated correctly and thus that we have a rather complete view of the object. 2024 PT5 is overall a quite red object, reflecting about twice as much incoming light at $2.35\mu{m}$ compared to $0.55\mu{m}$. The object has a strong $1.0\mu{m}$ absorption feature (with a band center around $0.94\pm0.01\mu{m}$, shorter than many but not all asteroids) and a relatively weak $2.0\mu{m}$ absorption feature, consistent with an overall rocky and silicate-rich composition. The spectrum of 2024 PT5 turns upwards beyond $\approx2.1\mu{m}$, which could be due to the surface of the object becoming more reflective at those wavelengths or due to excess thermal emission from the object if it had a particularly low albedo (see, e.g., \citealt{2005Icar..175..175R, 2009M&PS...44.1917R, 2022PSJ.....3..105K}). Silicate-rich surfaces are generally high enough albedo ($p_V > 0.1$) that thermal emission at these wavelengths is not substantial, but we return to this topic in the following section.

\section{Taxonomy} \label{sec:comp}
The reflectance spectrum of 2024 PT5 has all of the characteristics of a rocky surface with silicates on it, and therefore we see no reason to infer an artificial origin (e.g. spacecraft debris, lost rocket bodies, etc.). Furthermore, if the object had a high area-to-mass ratio like most space debris, it would have had a significant orbital drift consistent with radiation pressure (see again \citealt{2024PSJ.....5...96B}). We analyze whether or not this effect is found in the orbit in Section \ref{sec:orb}. The object thus very much appears natural in origin -- but where did it come from?

We applied two different analysis techniques to the combined visible/near-infrared reflectance spectrum of 2024 PT5 to diagnose its origin. First, we ran the spectrum through the online taxonomic classifier for the Bus-Demeo asteroid taxonomic system \citep{2009Icar..202..160D}. No asteroid taxonomic class fit PT5 well at all. The O- and Q-types aren't red enough, while the A-types (the reddest silicate-rich asteroid class) are too red. Other stony asteroid types, such as the S-types, are even worse fits primarily due to having the wrong spectral slope. PT5 is thus intermediate between several rocky asteroid classes in terms of spectral slope. In the recent large NEO spectral survey of \citet{2024PSJ.....5..131S}, those authors found a small number of NEOs with red slopes and weak $0.9-$ and $2.0-\mu{m}$ absorption features. Those authors suggested that their absorption features might be suppressed due to to shock darkened material (see, e.g., \citealt{2022PSJ.....3..226B}) or significant metal content. However, a metal-silicate mixture is likely a poor match for 2024 PT5 even if there are some qualitative similarities between the spectra of \citet{2024PSJ.....5..131S}'s ``Sx-types" and this new object (namely, the relatively weak $2.0-\mu{m}$ band). Increasing the amount of shock darkening or metal content on an asteroid makes its spectrum redder and its spectral features more muted or shallower at the same time -- this is inconsistent with 2024 PT5 being redder than any of the objects in the Sx-type sample \citep{2024PSJ.....5..131S} while still having a deep $0.9-\mu{m}$ band. If these processes were what was driving the red continuum slope of 2024 PT5, one would then expect that it would have even weaker spectral features than the objects in that sample. We thus conclude that shock darkening or high metal content are unlikely candidates to explain the reflectivity of 2024 PT5 as well.

As mentioned above, the band center of the object's one micron absorption feature is near to $0.94\pm0.01\mu{m}$ -- this is shorter than most asteroids whose centers are at slightly longer wavelengths. The exception to this generalization are the V-types, but those have much deeper $2.0-\mu{m}$ absorption features than PT5 does. Based on the band centers and relative depths, 2024 PT5's spectrum is pyroxene rich while the other comparably red asteroid spectra are olivine rich. In other words, even if some of the asteroid spectra can match certain aspects of PT5's spectrum, there is a significant compositional mismatch between it and the known asteroids.

As a test, we also fit the near-infrared ($0.85-2.45\mu{m}$) spectra separately and found somewhat better fits -- the Q-types being the best fit though still not sufficiently red -- but considering that our visible and near-infrared data are consistent (especially regarding the shape and center of the $1-\mu{m}$ band), we are inclined to believe the lack of good matches is real. While the object was observed at moderate ($\sim64^{\circ}$) phase angle, and thus some phase reddening should be expected \citep{2012Icar..220...36S}, phase reddening is not fully consistent with the spectrum obtained. Phase reddening (at least in Ordinary Chondrite-like powders, which we do not know a priori reflect light and phase redden like PT5) results in deeper $0.9-\mu{m}$ and $2.0-\mu{m}$ bands, inconsistent with our sample having band depth ratios similar to many Q-type asteroids as evidenced by the near-infrared-alone analysis. PT5 is too red at visible wavelengths to match a Q-type spectrum in the near-infrared, but is not red enough in the near-infrared to match an A-type asteroid overall. Finally, we note that none of the asteroid types that fit the rest of the spectrum well have a reflectivity increase beyond $\approx2.10\mu{m}$ at the level seen in our PT5 spectrum. As discussed above, a pyroxene-rich surface (or any rocky surface) is not expected to be dark enough to have significant thermal emission at these wavelengths at this heliocentric distance, so this is another indication that the known asteroid types are a poor fit for this object's reflectance spectrum. That said, the object is relatively small, so perhaps insights from the thermal states of larger asteroids do not apply well to $\approx10$-meter objects for which interal heat conduction might be more important. We thus conclude that while we cannot rule out 2024 PT5 having an origin in the asteroid belt from its reflectance spectrum (the challenges with putting asteroidal material onto such a near-Earth orbit aside), it would have to have a fundamentally rare kind of composition\footnote{For context, in the \citet{2019Icar..324...41B} sample, the single best fit type for 2024 PT5 -- `Qw', or a Q-type spectrum with an unusually red slope (denoted by the `w') -- represents about one third of one percent of the 1000$+$ asteroids studied.} among the Near-Earth Asteroids if that were the case.

\begin{figure}[ht!]
\plotone{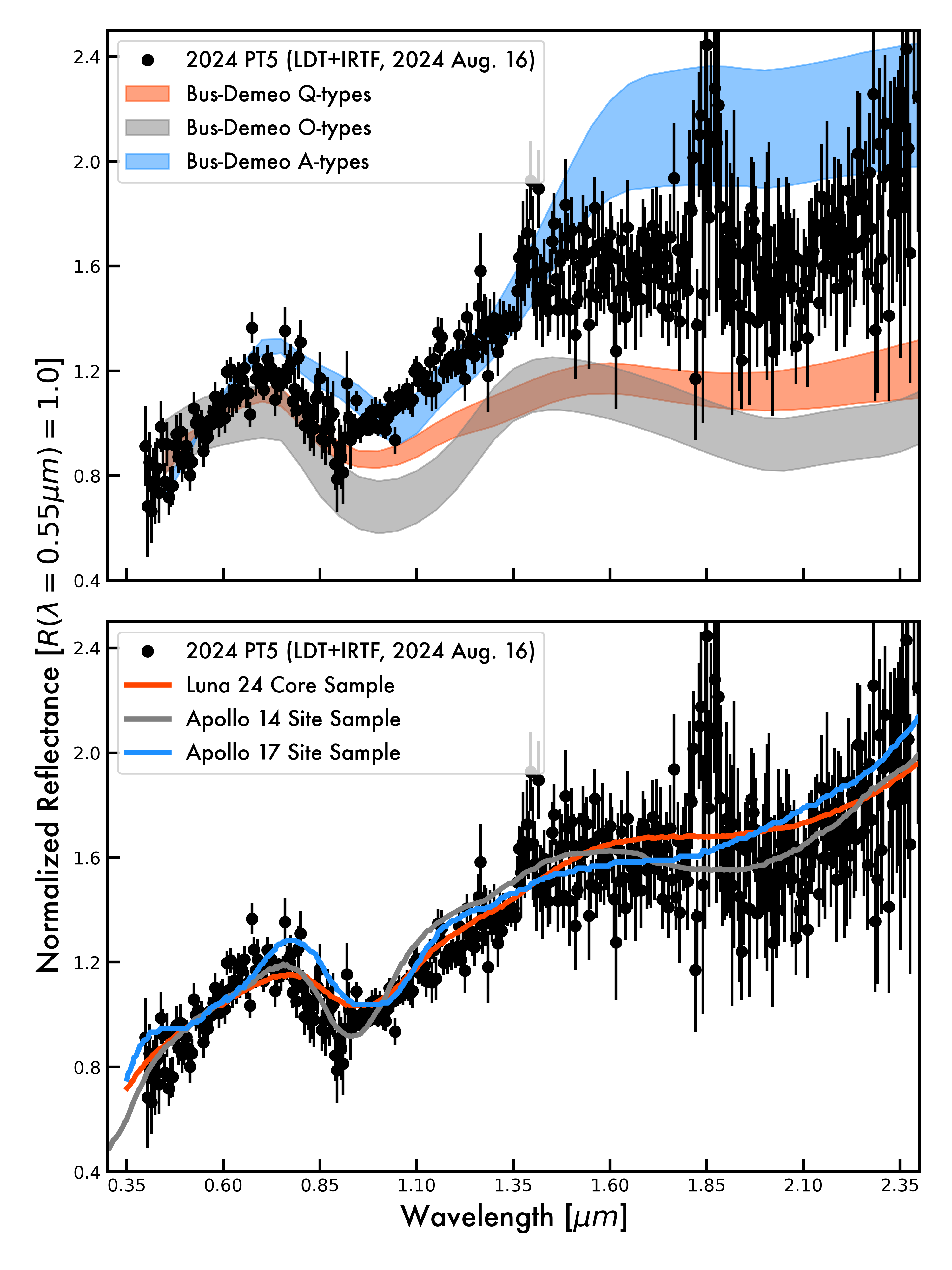}
\caption{The combined LDT+IRTF visible/near-infrared reflectance spectrum is shown (black) compared against relevant asteroid types from \citet{2009Icar..202..160D} (top panel) and returned samples from the Moon (bottom). As can be seen, none of the asteroid classes can fit the spectrum of 2024 PT5 well but the regolith samples from the Moon can fit all aspects of the spectrum relatively easily. The details of and limitations to these comparisons are described at length in the text.}
\label{fig:taxamoon}
\end{figure}

The second approach was to compare the combined reflectance spectrum of 2024 PT5 to every sample in the RELAB database \citep{1983JGR....88.9534P}, a collection of reflectance spectra for many kinds of materials -- from meteorites to terrestrial samples to lab-made materials. Given the challenges in fitting the telescopically-derived reflectance spectrum to any given asteroid type, we did not make any prefential cuts in albedo (e.g., if it had looked like a carbonaceous chondrite, we might have only compared to samples with bulk albedos less than $0.2$ or similar) of the samples studied and the only cut we made was to compare our spectrum to samples which had spectra which covered the full wavelength range ours covered. The three best fits samples are returned powders and mixed-grain-size samples from \textit{Luna 24}, \textit{Apollo 14}, and \textit{Apollo 17}. These three samples can be seen in Figure \ref{fig:taxamoon} compared to the combined telescopic spectrum of 2024 PT5. While there are still some minor mismatches between the laboratory spectra of these returned lunar samples and our reflectance spectra of 2024 PT5, the matches on the whole are radically better than the asteroid taxonomic types ($RMS_{lunar} / RMS_{Bus-Demeo} \approx 1/3$, with the best-fit lunar samples having spectral mismatches on the scale of typical error bars or better) and can fit the visible and near-infrared data simultaneously. We present the Band 1 centers and depths (derived following the methodology of \citealt{2020AJ....159..146S}) as well as visual ($0.4\mu{m}$ to $0.7\mu{m}$) and near-infrared ($0.85\mu{m}$ to $2.20\mu{m}$) spectral slopes (in units of $\%$ per $0.1\mu{m}$, following \citealt{1990Icar...86...69L}) as a quantitative summary in Table \ref{tab:values}.

\startlongtable
\begin{deluxetable*}{c|c|c|c|c}
\tablecaption{Spectral Properties of 2024 PT5 and Curve-Matched Lunar Samples.}
    \label{tab:values}
    \centering
    \startdata
        Object & Band 1 Center [$\mu{m}$] & Band 1 Depth [$\%$]& $S'_{VIS}$ [$\%/0.1\mu{m}$] & $S'_{NIR}$ [$\%/0.1\mu{m}$] \\
        \hline
         2024 PT5 & $0.94\pm0.01$ & $27\pm1$ & $16\pm1$ & $4.7\pm0.2$\\
         Luna 24 & $0.98\pm0.01$ & $19\pm2$ & $10.2\pm0.1$ & $3.98\pm0.05$\\
         Apollo 14 & $0.94\pm0.01$ & $28\pm1$ & $11.9\pm0.2$ & $3.6\pm0.1$\\
         Apollo 17 & $0.99\pm0.01$ & $26\pm1$ & $8.4\pm0.2$ & $3.93\pm0.05$
    \enddata
\end{deluxetable*}

One compelling detail is how well the lunar samples can fit the long wavelength ($>2.1\mu{m}$) reflectivity of PT5. If 2024 PT5 really is made of lunar material and these matches are not spurious or otherwise challenging to interpret, the best-fit samples have albedos between $0.13 < p_V < 0.17$, resulting in a diameter of $8$ m $< D < 12$ m for 2024 PT5. We tried several consistency checks, from performing these analyses with the data unbinned or binned to courser resolutions to re-reducing the data with different extraction apertures and found no meaningful differences in our results. 2024 PT5 reflects light far more like material from the Moon than it does relative to almost any known asteroid.

Should 2024 PT5 actually be composed of lunar material, the question of its size becomes critical to interpreting any spectral matches. In particular, smaller asteroids with less surface gravity -- and especially those in rapid or changing spin states -- would be expected to have a harder time holding on to fine regolith on their surfaces. While much of this expectation is on theoretical grounds due to the challenges in characterizing such small objects, at least some small asteroids seen to burn up in the atmosphere appear to have no regolith on their surfaces at all \citep{2024PSJ.....5..253K}. The three best-matched lunar samples above are all of relatively fine powders. How meaningful, then, is it to match an 8-to-12 meter asteroid expected to have little or no regolith with a laboratory spectrum of powders? For Moon-derived materials, we have additional ways to assess this question compared to traditional asteroid surfaces. A visible/near-infrared spectrum obtained by the Chang’E-4 lander's rover Yutu-2 of a rock tens of centimeters across \citep{10.1093/nsr/nwz183} found it to be comparably red to our asteroid. Those authors argued that despite the rock appearing to be regolith-free, it may have been coated in a thin dusting of fine powder -- in any case, the spectra of fine powders is clearly still relevant to understanding the spectral behaviors of larger surfaces. Even if we are not certain what kind of grain size is the most appropriate to attempt to match our spectra to, there are rocks on the Moon which have similar spectral behaviors to 2024 PT5. We return to the topic of grain size in Section \ref{sec:disc}.

\section{Orbital Assessment and Context} \label{sec:orb}

Asteroid close encounters as slow as that of 2024 PT5 are very rare. We inspected JPL's CNEOS Close Approach Data, which records known close approaches of objects in JPL's Small-Body Database\footnote{Available for query at: \url{https://ssd-api.jpl.nasa.gov/doc/cad.html}.}. We found that in this century there are only 9 other asteroids with flybys slower in $v_\infty$ than 0.7 km/s and 3 objects with negative energy during closest approach: 2022 NX1 \citep{2023A&A...670L..10D}, 2020 CD3 \citep{2021ApJ...913L...6N} and 2006 RH120 \citep{2009A&A...495..967K}. These are the only ones found among $\sim$17,000 recorded close approaches between 2000 January 1 and 2024 November 15. The typical $v_\infty$ range is 3 to 22 km/s (2.5 and 97.5 percentiles). 

Such slow encounter may suggest an artificial origin of 2024 PT5, a hypothesis we can investigate from the object's motion. Artificial objects are much lighter than asteroids, and thus solar radiation significantly perturbs their motion. A heliocentric inclination of $\sim$1 deg is common among artificial objects, so it is not a strong argument against the artificial origin of  2024 PT5 as it was for 2003 YN103 \citep{2004M&PS...39.1251C}. This effect provides a way to distinguish between natural and artificial objects \citep{2024PSJ.....5...96B}. This non-gravitational acceleration can be parametrized by the area-to-mass ($A/m$) ratio. Typical $A/m$ values for artificial objects can be in the order of $\sim0.01$ m$^2$/kg. For example, space debris WT1190F has an estimated $A/m=0.0118\pm0.0005$ m$^2$/kg \citep{doi:10.2514/6.2016-0999}, or J002E3, thought to be the upper stage of Apollo 12's Saturn V\footnote{JPL Press release of 2002 Sept 20: \url{https://www.jpl.nasa.gov/news/first-confirmed-capture-into-earth-orbit-is-likely-apollo-rocket/}. Orbital model from NASA/JPL Horizons On-Line Ephemeris System, \url{https://ssd.jpl.nasa.gov/horizons/}, data retrieved 2024 November 15.}, whose orbit is modeled with $A/m$ = 0.0079  m$^2$/kg in NASA/JPL's HORIZONS System \citep{2001DPS....33.5813G}. $A/m$ values for asteroids are typically two orders of magnitude smaller \citep[found $(2.97\pm0.33)\cdot10^{-4}$, $(3.35\pm0.28)\cdot10^{-4}$ and $3.2\cdot10^{-4}$ m$^2$/kg for 2009 BD, 2012 LA and 2011 MD, respectively]{2012NewA...17..446M,2013Icar..226..251M,2014ApJ...788L...1M}. We attempted to estimate $A/m$ for 2024 PT5 from its astrometry, and find an estimate of $0.0\pm1.3\cdot10^{-4}$ m$^2$/kg, which rules out solar radiation pressure at the magnitudes expected for artificial objects but is compatible with what is expected for a natural one. 

\begin{figure}[ht!]
    \centering
    \includegraphics{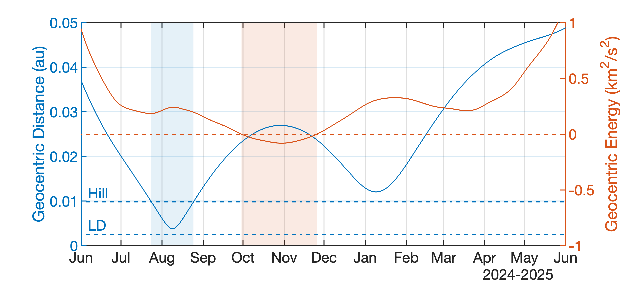}
    \caption{Geocentric distance and energy of 2024 PT5 between June 2024 and June 2025. The times when the distance $<$ Hill distance and energy $<0$ are highlighted by the shadowed areas, showing that these conditions do not occur simultaneously.}
    \label{fig:1d_2024}
\end{figure}

The 2024 encounter of 2024 PT5 with Earth was so slow that the geocentric distance stayed within 0.03 au for months, from June 2024 to February 2025. We show the geocentric distance and energy in Figure \ref{fig:1d_2024}. When 2024 PT5 was inside Earth's Hill sphere (2024 July 24 to 2024 August 25) the geocentric energy remained positive. Similarly, when 2024 PT5's geocentric energy was negative, the geocentric distance was much larger than the Hill sphere radius. This means that the acceleration from Sun's gravity dominated over Earth's, and thus 2024 PT5 was not captured by the Earth. We discuss commonly used quasi-satellite terminology in Section \ref{sec:termi}. Nonetheless, the encounter is of particular interest because 2024 PT5 is currently transitioning from the Aten class (a$<$1 au, Q$>$0.983 au) to Amor (a$>$1 au, q$>$1.017 au). This transition describes a horseshoe trajectory, from a faster orbit when a$<$1 au to slower orbit after its close approach on 2024 August 8. 2003 YN107 is another example of such horseshoe transitions prior to its discovery encounter \citep{2004M&PS...39.1251C} as well as 2022 NX1 \citep{2023A&A...670L..10D} during its 2022 encounter.

\begin{figure}[ht!]
    \centering
    \includegraphics[scale=1]{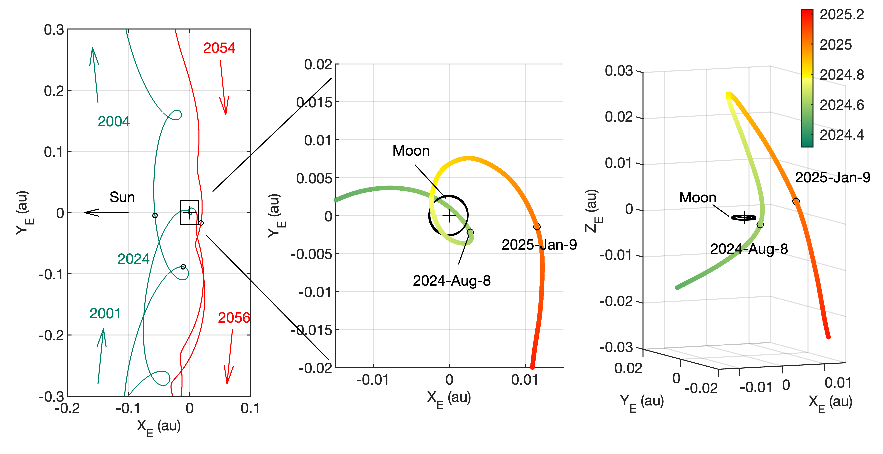}
    \caption{Geocentric orbit of 2024 PT5 in Earth-ecliptic rotating frame (Sun to -X$_E$ direction). The left panel includes the 2002-2003 and 2055 close approaches (circled, details in Table \ref{tab:CAT}) and the 2024-2025 horseshoe transition. The center and right panels show in detail the 2024-2025 close approaches in the same frame, including the out-of-ecliptic component}
    \label{fig:3d_2024}
\end{figure}

\begin{figure}[ht!]
    \centering
    \includegraphics[scale=1]{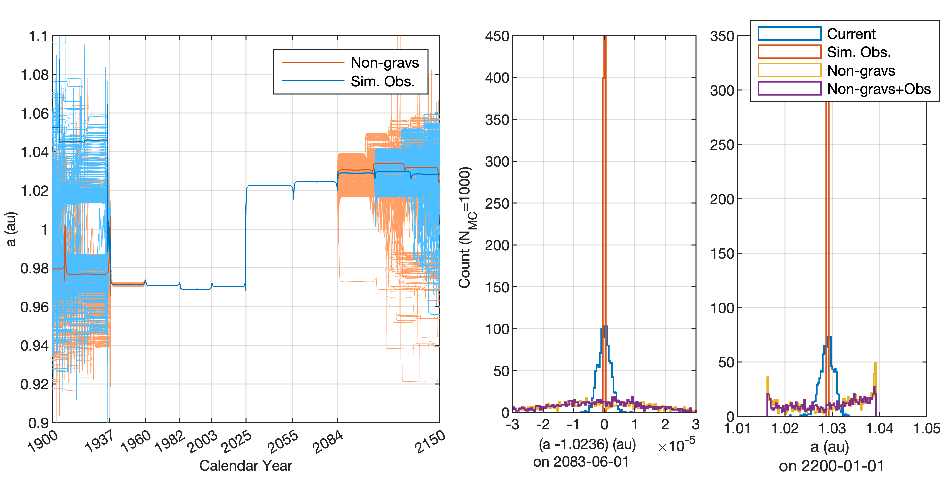}
    \caption{Heliocentric semi-major axis of 2024 PT5 between years 1900 and 2150. Right 2 panels show histograms of the distribution before and after the 2084 encounters using 4 solutions: Current, gravity-only solution; Sim. Obs, including the additional simulated observations; Non-gravs, solution with gravity and non-gravitational accelerations; Non-gravs+Obs, including both non-gravitational accelerations and the additional simulated observations. The semi-major axis distribution in presence of non-gravitational accelerations becomes non-Gaussian due to the wider distribution in timing in the first 2084 encounter ($\sigma$-$t_{CA}=0.94$ h, $v_\infty=0.766$ kms$^{-1}$) and complex interactions of the slow encounter.}
    \label{fig:ae_pm200yr}
\end{figure}

Figure \ref{fig:3d_2024} shows the position of 2024 PT5 relative to the Earth in an Earth-ecliptic rotating frame such that the $-X_E$ direction points to the Sun at all times and the $+Z_E$ direction is parallel to Earth's orbital angular momentum. Before 2024, 2024 PT5 has a faster orbital velocity hence it moves in the $+Y_E$ direction during the 2003 encounters portrayed in Figure \ref{fig:3d_2024}. After 2024, the Earth is faster and 2024 PT5 moves in the $-Y_E$ direction. The transition between the two motion regimes is explained by the transition between $a<1$ and $a>1$ au, which is more explicit in the evolution of the semi-major axis of Figure \ref{fig:ae_pm200yr}.

Figure \ref{fig:3d_2024} also shows the orbital geometry of the discovery of 2024 PT5. 2024 PT5 approached the Earth from close to the Sun direction, and it was discovered slightly before its closest approach at 0.00379 au (1.47 LD) on 2024 Aug 8. This approach was hyperbolic: its geocentric energy was and remains positive until September 29, when it is much further from the Earth at 0.023 au (9 LD). Despite the low heliocentric pre-encounter inclination (0.96 deg), the flyby had a significant out-of-plane component as shown in the third panel of Figure \ref{fig:3d_2024}, coming in from the ecliptic $-Z_E$ direction to reach a $+Z_E$ maximum. Afterwards, 2024 PT5 continues to move in the anti-Sun ($+X_E$) and back into $-Z_E$, where it has its second closest approach of 0.01204 au on Jan 9 2025 (4.68 deg), providing favorable conditions for follow-up observations. The post-encounter heliocentric inclination becomes 1.71 deg.

Table \ref{tab:CAT} shows the close approaches of 2024 PT5 with the Earth for as long as the uncertainty in the orbital position is small enough to reliably determine further close approach circumstances. Before the 2024 close approach with Earth, 2024 PT5 was experiencing Earth encounters roughly every 20 years, of asymptotic relative velocity between 1 and 3 km/s depending on the phase of the oscillation around the horseshoe orbit. After this encounter, the encounter frequency is close to 30 years. Due to the very strong interactions with the Earth-Moon system, the trajectory becomes stochastic beyond the encounters of 1937 and 2084.  It is possible that the outcome of the 2084 and 1937 encounters becomes deterministic after further observations. Throughout the current apparition additional astrometric observations are being made available, which reduce the uncertainty of the orbit. We simulated weekly astrometry at $0.5"$ of precision in combination with a radar delay observation on 2025 January 9 of 1 $\mu s$ precision. As for the dynamical model, we included the gravitational accelerations of the planets, Moon, and 16 largest main belt asteroids using ephemeris models DE441 \citep{Park2021} and SB441-N16 \citep{Farnocchia2021sb441}. In addition, objects the size of 2024 PT5 are expected to experience non-gravitational accelerations. We include these by introducing $A_1$ and $A_2$ accelerations in the radial-transverse directions, which correspond respectively to accelerations due to solar radiation pressure and the Yarkovsky effect \citep{2015aste.book..815F}. We model these accelerations with Gaussian distributions of $A_1 \sim N(5,2.5)\cdot10^{-11}$ au/d$^2$ and $A_2 \sim N(0,2)\cdot10^{-12}$ au/d$^2$. This approach leaves us with 4 sets of solutions: gravity only, gravity only with simulated observations, gravity with non-gravitational accelerations, and including both non-gravs and simulated observations. When including these effects the uncertainty in semi-major axis increases and the orbit becomes stochastic sooner. Table \ref{tab:CAT} shows the increase in uncertainty of close approach dates and distances using the non-gravitational acceleration model, as well as differences in the nominal dates and distances.

We propagated 1000 trajectories from the previous solutions in Monte Carlo fashion in Figure \ref{fig:ae_pm200yr}. The semi-major axis evolution over time is shown for the model with largest uncertainty and the model with the smallest: the non-gravitational accelerations model in orange, and the gravity-only solution with simulated observations in blue. The uncertainty evolves linearly until 2084. The second panel shows how the distribution of particles remains Gaussian and roughly around the same mean in all models. The non-gravitational solution is actually slighly off the nominal, as seen by the small differences in the nominal dates of closest approach in Table \ref{tab:CAT}. After the 2084 encounter, the range of possible outcomes is very diverse. The gravity-only solution with simulated observations best constrains the post-encounter orbit, but even in that case the next encounters are already stochastic. Most remarkably, the non-gravitational range of values spreads the encounter conditions in 2084 so that the range of possible post-encounter orbits is wide, and remarkably non-Gaussian. This is explained by an initially large uncertainty in timing: the close approach date has an uncertainty of 0.94 h, which translates into a large position error considering $v_\infty=0.766$ km/s. To understand how the uncertainty spreads along the orbit we can map it into position using the encounter B-plane \citep{2003A&A...408.1179V,2019CeMDA.131...36F}. In the \"{O}pik B-plane, the $\xi$ coordinate is associated to the MOID (Minimum Orbit Intersection Distance), which corresponds to the closest possible approach distance. The $\zeta$ coordinate represents the relative timing before or after the closest possible encounter. For the 2084 encounter, $\zeta=(8.5\pm5.0)\cdot 10^5$ km, and $\xi=(-21.25\pm0.17)\cdot 10^5$ km (Uncertainties are 1-$\sigma$ standard deviation from Monte Carlo). This decomposition serves to illustrate how most of the position uncertainty is in timing approaching the Earth, although at such slow velocities defining the B-plane can be inaccurate \citep{2015CeMDA.123..151V}. After the first 2084 encounter, 2024 PT5 will stay close to the Earth for months until a second close approach. The two encounters are a complex non-linear interaction that causes an initially large uncertainty to become non-Gaussian. The gravity-only solution with simulated observations produces a deterministic outcome of the 2084 encounters. But even then, the orbit becomes stochastic at subsequent encounters. 2024 PT5 is in a naturally stochastic orbit that does not allow a deterministic prediction of its origin or future beyond 100 years. This is common among some NEAs, but more extreme in objects of Earth-like orbits.

\begin{table}
    \centering
    \caption{Earth close approaches (CA) table. Uncertainty in $t_{CA}$ is 1-$\sigma$, whereas uncertainty in $d_{CA}$ corresponds to half of the maximum-minimum $d_{CA}$. Table generated using the observational arc ranging 2024-08-07 - 2024-10-24, for solutions with Gravity Only (GO) or using the stochastic non-gravitational acceleration model (NG). The close approaches of 1937 and mid-2084 are poorly defined in the NG model.}
    \label{tab:CAT}
    \begin{tabular}{rcc|cc|cc}
$t_{CA}$ Date& Day & Day & $d_{CA}$ (au)& $d_{CA}$ (au)& $v_{\infty}$ & (kms$^{-1}$)  \\
 & GO & NG & GO & NG & GO & NG  \\\hline 
1937-Oct&9.176 $\pm$ 0.049 &  $\sigma>1$ & 0.016 $\pm$ 1.8e-02 & $\sigma>0.1$ & 1.056 & - \\
1960-Mar&18.078 $\pm$ 0.012 & 16.409 $\pm$ 0.099 & 0.056 $\pm$ 2.8e-04 & 0.056 $\pm$ 2.0e-03 & 1.390 & 1.402 \\
1960-Oct&9.041 $\pm$ 0.001 & 8.911 $\pm$ 0.008 & 0.099 $\pm$ 3.7e-03 & 0.102 $\pm$ 3.0e-02 & 3.566 & 3.649 \\
1982-Apr&2.163 $\pm$ 0.001 & 2.326 $\pm$ 0.009 & 0.056 $\pm$ 6.6e-05 & 0.056 $\pm$ 5.0e-04 & 1.333 & 1.332 \\
1982-Oct&7.154 ($\sigma<10^{-3}$) & 7.163 $\pm$ 0.001 & 0.080 $\pm$ 3.9e-04 & 0.080 $\pm$ 3.0e-03 & 3.022 & 3.013 \\
2002-Jul&29.559 ($\sigma<10^{-3}$) & 29.560 ($\sigma<10^{-3}$) & 0.090 $\pm$ 5.2e-05 & 0.090 $\pm$ 5.4e-04 & 3.288 & 3.287 \\
2003-Feb&11.208 ($\sigma<10^{-3}$) & 11.181 $\pm$ 0.002 & 0.057 $\pm$ 4.0e-06 & 0.057 $\pm$ 4.1e-05 & 1.432 & 1.431 \\
2024-Aug&8.833 ($\sigma<10^{-3}$) & 8.833 ($\sigma<10^{-3}$) & 0.004 ($\sigma<10^{-6}$) & 0.004 ($\sigma<10^{-6}$) & 0.694 & 0.694 \\
2025-Jan&9.091 ($\sigma<10^{-3}$) & 9.090 ($\sigma<10^{-3}$) & 0.012 ($\sigma<10^{-6}$) & 0.012 ($\sigma<10^{-6}$) & 0.780 & 0.780 \\
2055-Nov&8.720 $\pm$ 0.001 & 8.797 $\pm$ 0.013 & 0.036 $\pm$ 9.0e-06 & 0.036 $\pm$ 9.1e-05 & 0.543 & 0.544 \\
2084-Jan&7.285 $\pm$ 0.005 & 6.974 $\pm$ 0.039 & 0.011 $\pm$ 3.9e-04 & 0.012 $\pm$ 5.2e-03 & 0.767 & 0.766 \\
2084-May&8.058 $\pm$ 0.053 &  $\sigma>1$ & 0.022 $\pm$ 2.6e-03 & $\sigma>0.1$ & 0.754 & - \\
    \end{tabular}
\end{table}

\section{Discussion} \label{sec:disc}
In this manuscript, we have presented telescopically derived reflectance spectra of the newly discovered near-Earth object 2024 PT5 and argued that it reflect light more like lunar samples than almost any other asteroid type. We have also discussed the specifics of this objects orbit, including both the challenges in getting asteroidal material onto such low $v_\infty$ orbits and the past close encounters this object may have had with the earth. In this Section, we discuss a few outstanding implications and open questions that arise from finding what is likely the second NEO that originated as ejecta from a significant impact on the Moon.

\subsection{Two Different lunar-like NEOs}
Perhaps the most significant conclusion to finding a second near-Earth object with an apparently Moon-like surface composition is the realization of lunar Ejecta as a genuine population of objects. The Quasi-Satellite Kamo`oalewa \citep{2021ComEE...2..231S} has a slightly redder spectrum than 2024 PT5, but the higher quality of our data at longer wavelengths (the Quasi-Satellite was significantly dimmer, so only photometry was obtained beyond $\approx1.25\mu{m}$) makes a discussion of how different the two spectra are only qualitative. At the very least, the two lunar NEOs do not look identical. \citet{2021ComEE...2..231S} argued that the red spectrum of Kamo`oalewa was partially due to space weathering -- an exposure time of a few million years was likely sufficient to explain its surface properties and was similar to its approximate dynamical lifetime and even the age of the crater that \citet{2024NatAs...8..819J} suggested it came from, Giordano Bruno. If correct, perhaps 2024 PT5 has a somewhat younger surface than the larger Kamo`oalewa. In any case, PT5 is smaller than Kamo`oalewa and thus the craters that are energetic enough to produce an object its size are more common -- a more recent ejection age, and thus a `younger' surface might be preferred from that argument as well. (Granted, smaller fragments would be more common than larger ones in cratering events of any size as well.) Further work to study these two objects and to find more lunar-like NEOs (see next subsection) will be needed to ascertain the origin of these differences and how they can be related to the circumstances of their creation. At any rate, the smaller size of PT5 means that we are approaching being able to study the impactors and outcomes from the kinds of small impacts seen regularly by the Lunar Reconaissance Orbiter \citep{2016Natur.538..215S}.

As described in Section \ref{sec:intro}, significant work has been done since Kamo`oalewa was proposed to be lunar ejecta in trying to determine what can be learned about the specifics of how impacts work on the Moon and to see what lunar craters might be likely culprits for its ejection. The three best spectral matches to our combined visible/near-infrared reflectance spectrum are from a core sample from Luna 24 (sampled the Mare) and site samples from Apollo 14 (sampled the highlands) and Apollo 17 (sampled both kinds of terrains). While this might suggest that we have no ability to distinguish between origins in the Mare or the highlands for PT5, an inspection of Fig. \ref{fig:taxamoon} offers a few insights. First, in terms of band centers and depths, the highlands material from Apollo 14 clearly matches better -- but that same sample fares slightly worse along the continuum points far from the two primary bands \textbf{(see again Table \ref{tab:values})}. This is most pronounced in the region from $1.10-1.35\mu{m}$, where PT5's reflectance is lower than what is predicted by all but the Luna 24 sample of the Mare. Other than at those wavelengths, the Apollo 14 performs as well as the others -- in that they all more or less go through the object's 1-$\sigma$ error bars. Given that differences in average grain size, phase angle, and space weathering state, can affect the reflectance continuum more easily than the band centers, we argue for a light preference for Highlands material over the Mare, but future observations with higher SNR could better resolve the shape of PT5's bands and thus more clearly link it with specific terrains. In particular, finding ways to measure the albedos of these lunar-derived NEOs (likely a task for JWST due to their faintness) would help break some of the grain size degeneracies brought up in Section \ref{sec:comp}.

\subsubsection{Other and Future Observations}
As mentioned in Section \ref{sec:obs}, \citet{2024arXiv241108029B} also obtained photometric observations of 2024 PT5 -- though their results are rather different than ours. While their colors for PT5 are similar but not identical to ours (they have a $1-\mu{m}$ absorption feature, indicating the object has a rocky composition), their lightcurve is wholly different. While our observations showed no obvious sign of variability greater than a fraction of a tenth of a magnitude over approximately an hour, their sparser observations showed a substantial $0.3-$mag variation over the same timescale. It seems likely that much of the differences in our lightcurves stem from the fact that \citet{2024arXiv241108029B} obtained their observations more than a month after ours when the object was more than five magnitudes, or a factor of a hundred times, fainter. That said, it is possible the object is in a tumbling rotational state and thus both lightcurves could be reconciled. Future observations to constrain the rotational state and albedo of this object, as well as to constrain the phase-reddening behavior of its surface, are highly encouraged.

\subsection{Hunting for More Moon Rocks}
If there really is a population of Moon Rocks out there waiting to be discovered on near-Earth orbits, they almost certainly are rare members of the NEO population. We can conservatively estimate the number of lunar  fragments in the known NEO population by considering the fraction of NEOs that have close encounters with the Earth inside of 1 lunar distance (3.4\% of the overall NEO population), and the fraction of those that have $v_\infty < 2.4$ km/s (1.3\% of the close encounter population). These coarse statistics suggest $\sim16$ NEOs in the current catalog could be lunar ejecta. This assumption regarding encounters inside of a lunar distance is not unreasonable given the small sizes of temporarily captured objects and the strong discovery biases against such small objects at greater geocentric distances. This estimate however is almost certainly overly conservative. \citet{2023ComEE...4..372C} showed that $<10$\% of lunar ejecta end up in states co-orbital with the Earth, while the other 90\% either re-impact the Moon, impact the Earth, or evolve into typical Aten or Apollo NEO orbits. If the evolution of lunar ejecta into Aten or Apollo orbits is this efficient, there could 5-10$\times$ more pieces of the Moon in the current NEO catalog than our conservative estimate. In other words there could be upwards of $\sim100$ known NEOs that originated on the Moon. 

Regardless of the size of the lunar NEO population, how can we distinguish them from the backround population? If Kamo`oalewa and 2024 PT5 are broadly representative of the lunar NEO population, then we might expect there to be an overabundance of redder-than-typical silicate rich asteroids on very near-Earth orbits. Formally, 2024 PT5 would be classified as a Qw-type asteroid in the \citet{2009Icar..202..160D} scheme -- but these represent less than half a percent of the surveyed NEO population \citep{2019Icar..324...41B}. This leaves three non-exclusive possibilities: the lunar NEO population is fundamentally small in number, the population is dominated by objects at small enough sizes that current surveys are inefficient at finding and characterizing them, and/or the other lunar NEOs have different reflective properties.

Surveys designed to find more objects in Quasi-Satellite, Horseshoe, and near-Horseshoe orbits will help to assess the overall size of the population. The Vera Rubin Observatory's LSST might assist in finding the smaller members of this population due to its fainter limiting magnitude -- especially useful if most of them are small in size -- but work will need to be done to assess if the planned survey cadence will be able to find these kind of objects and what kind of follow-up will be needed to confirm them.

The size distribution of lunar NEOs could be assessed in two ways. The first would simply be to find more objects that are of likely lunar origin and then to directly calculate their size distribution -- but this could take years of surveying, and then would still be affected by observational biases associated with surveying for a population of objects with an unknown orbital distribution. Further work to model the distribution of lunar ejecta such that a constraining survey could be performed, or at the very least such that promising objects could be identified more easily, is clearly of use. Another way would be to inspect the size distribution of coherent fragments released in simulations of impacts of various sizes into the lunar surface (see again \citealt{2024NatAs...8..819J}) and then to weight the derived population by how frequent impacts of a given size are and what fraction of ejecta in each scenario reaches the NEO population.

To assess the third scenario, whereby other lunar NEOs have different surface reflectivities than the two known, we took 863 reflectance spectra of lunar materials over the visible/near-infrared wavelength range and applied an RMS-minimization fit to the asteroid types in \citet{2009Icar..202..160D}. Of these spectra, only forty five `matched' one of the Bus-DeMeo classes with an RMS value less than $0.2$ -- and the best comparisons only had a slightly better RMS of $\approx0.16$. Upon close inspection, all of the lunar spectra studied have features that would clearly make them an outlier in the classes they are closest to -- the D- and B-types have absorption features, the A-types have insufficiently deep bands, the X- and K-types have bands that are too deep, and so on. Should one only acquire (or be able to acquire) visible wavelength observations, however, the fraction of lunar spectra that fit within one of the Bus-Demeo classes jumps considerably ($\geq60\%$) -- lunar samples show a wide range of visible slopes and a wide range of depths and centers for their $1.0-\mu{m}$ features. Near-infrared data alone of lunar material has a lower but still significant fraction of objects ($\geq30\%$) which can be well fit by a Bus-Demeo class. Only through combining visible and near-infrared data can lunar material be confidently identified as different from asteroidal material, and even then it requires at least moderate SNR. A deeper dive into this analysis, including Figure \ref{fig:taxamoon} which shows some of the `matched' lunar spectra, is available in the Appendix.

Based on what has been discussed in this section and in other parts of this paper, we summarize how we have approached identifying lunar material in a given NEO into the following checklist. Should the object meet a majority of the criteria, then a lunar origin is quite possible and further investigation is warranted (e.g., a more advanced dynamical investigation). 

\textbf{Orbit: } Does the object have an orbit that is very similar to the Earth-Moon system, specifically with a low $v_\infty$ value? Lower values are required for recently ejected objects, but preferred for the whole population including the dynamically evolved component. Is the object on, or has the object been on, a horseshoe or co-orbital orbit/trajectory around or near the Earth?

\textbf{Size: } Is the object less than a few hundred meters in size? Smaller is preferred, as cratering events produce more small pieces of ejecta than large ones.

\textbf{Surface: } Does the object's reflectance spectrum match poorly to the known asteroid types? $RMS > 0.16$ is required, higher values are preferred. Does the object show a $1-\mu{m}$ band with a relatively short ($<0.93\mu{m}$) band center? Are there returned lunar samples or lunar meteorites which can match the reflectance of the object significantly better than any of the studied asteroid types?

For 2024 PT5, the answers to each of these questions is positive. A few more comments and a summarizing flow chart (Figure \ref{fig:app_lunar_flowchart}) of this kind of decision tree, including the processes by which artificial objects may be identified and removed, is also available in the Appendix.

\subsection{Asteroid Source Models and Planetary Defense}
In the past two decades, the source regions of near-Earth asteroids have been increasingly scrutinized and understood through modeling their source regions and the dynamical pathways that bring them into near-Earth orbits (see, e.g., \citealt{2002Icar..156..399B, 2018Icar..312..181G, 2024Icar..41716110N}). As discussed in Section \ref{sec:intro}, these models have tremendous utility from assessing impact hazards to inferring properties of newly discovered asteroids to tracing back individual meteorites and fireballs to the Main Belt. At present, these models do not include lunar ejecta as a source of near-Earth asteroids. At the very least, the discovery of a new source of near-Earth objects could lead to significant refinements in each of these models -- and thus to the estimated origins of many other objects on particularly near-Earth orbits.

While these models have a variety of use cases outlined in Section \ref{sec:intro}, one particular aspect to which the discovery of a population of lunar NEOs would be relevant would be planetary defense. Impacts onto the Moon naturally produce ejecta on very near-Earth orbits, and much of it ends up re-impacting the Earth and Moon afterwards (see again \citealt{2023ComEE...4..372C}) as opposed to reaching long-term stable orbits. Future work will be needed to assess the risk these kinds of objects pose towards the Earth.

\subsection{Terminology for Earth-orbit-like asteroids}\label{sec:termi}

Objects that fly by the Earth so slowly remain at small distances for extended period of times. On occasion, they can complete full revolutions around the Earth or their geocentric energy can become negative for periods of time. For these reasons, they have been classified as mini-moons, temporarily captured orbiter, satellite, or fly-by asteroid \citep{2018FrASS...5...13J}. The quasi-satellite term is also used for objects that remain close to the Earth such as (469219) Kamo`oalewa, even if they may not experience Earth close approaches and stay outside of Earth's sphere of influence. \citet{2017AJ....153..155H} provide analytical insight in the conditions for capture given an eccentric planet. Examples of objects fulfilling these conditions are 2020 CD3 or 2006 RH120, which orbited around the Earth for an extended period of time \citep{2009A&A...495..967K,2021ApJ...913L...6N}. 2022 NX1 had negative geocentric energy during its encounter although it did not complete a revolution around the Earth \citep{2023A&A...670L..10D}.

For an object to be captured, it should be sufficiently close to the Earth while having negative geocentric energy. For the distance threshold, the Hill sphere provides a meaningful threshold considering that it corresponds to the $L_1$ and $L_2$ equilibrium points. Previous works suggested the use of 3 Hill spheres \citep{1996Icar..121..207K}, although without a dynamical justification for such threshold. Being inside the Hill sphere of the Earth seems a sensible requirement as it ensures that Earth's gravity dominates an object's motion. Therefore, we require that the object is no further than one Hill sphere radius and that the energy is negative for an object to be considered as captured. These conditions are not simultaneously met by 2024 PT5 during its 2024 close approaches. Its motion is still remarkably close to that of Earth, but it is dominantly heliocentric.


\section{Conclusions}
2024 PT5 is a recently discovered near-Earth object (NEO) which spent months in late 2024-early 2025 in the immediate vicinity of the Earth. The origin of these very near-Earth objects is a topic with widespread interest, both because current asteroid source region models do not predict their existence in any significant number \citep{2018Icar..312..181G, 2024Icar..41716110N} and thus they may be sources not yet recognized or considered by those models. In particular, the potential source most relevant to this study is the Earth's Moon. The Earth Quasi-Satellite (469219) Kamo`oalewa has been suggested to be ejecta from an impact onto the surface of the Moon based on spectroscopic evidence \citep{2021ComEE...2..231S}, later supported by dynamical \citep{2023ComEE...4..372C} and cratering \citep{2024NatAs...8..819J} simulations. If a population of NEOs from the Moon does exist in number, this would provide a way to link lunar and NEO science, both by studying the specifics of crater formation (what kind of materials are ejected in what states from what depths) and through refining near-Earth object source region models to the benefit of both those working on planetary defense and those attempting to age-date the lunar surface. In this paper, we have presented a comprehensive telescopic and dynamical investigation into the properties, trajectory, and origin of 2024 PT5 and reached the following conclusions:

The surface of 2024 PT5 reflects light much more like lunar materials than it does any known asteroid type, much like Kamo`oalewa. This is the case both based on a bulk RMS analysis (e.g., what is the RMS difference between PT5 and the comparison spectrum as assessed over the whole visible/near-infrared wavelength range studied) and in terms of matching the shape and depth of its spectral bands. Compositionally, the spectrum of 2024 PT5 appears pyroxene rich -- asteroids with similar spectral slopes are generally olivine-rich, and thus fundamentally different. On the basis of the centers and depths of the spectral bands, an origin in the lunar highlands may be preferred. Assuming that the albedo of lunar samples that best matched PT5 are applicable, we estimate a diameter for the object between 8 and 12 m. We saw a flat lightcurve in our observations on 2024 Aug. 14, indicating that the object either was roughly spherical along the line of sight at that time or it has a rotation period shorter than a minute or longer than an hour.

During the 2024 close approach to Earth, 2024 PT5 transitioned from an Aten orbit to an Apollo one while moving along its horseshoe trajectory. 2024 PT5 has had significant close encounters with the Earth during the past few decades, but nothing as slow as that in August 2024. Asteroid close encounters as slow as 2024 PT5's are rare, especially so for those with negative binding energies -- beyond PT5, only three objects since the year 2000 have had similar encounters. At present, numerical integration of 2024 PT5's orbit becomes non-deterministic before 1937 and after 2084, though future astrometric characterization will assist in this.

The discovery of a second NEO derived from the Moon strongly suggests that there is a population waiting to be recognized, and a ballpark estimate suggests that there may be as many as sixteen NEOs currently known which are sourced from the Moon. NEO source region models will need to be refined, and a clearer understanding of this population will be highly useful to quantify planetary impact hazards on the Earth and to date the recent history of the Moon. We discuss several ways forward to find more of and characterize this population. Notably, we found through comparison with laboratory spectra of lunar samples that identification of an object as lunar in origin requires both visible and near-infrared data to determine with any confidence. While visible and near-infrared data on their own can frequently be matched to either asteroidal types (e.g. from the Bus-Demeo system, \citealt{2009Icar..202..160D}) or lunar samples, a combination of both visible and near-infrared data allows for significantly higher confidence in an assignment of an object in this kind of orbit as lunar or asteroidal. Given the small sizes expected for this population, large telescopes and new techniques may be necessary to acquire the required observations. We thus developed a checklist for future observers to refer to as they attempt to understand the properties of newly discovered objects on very Earth-like orbits. Lastly, we also discuss terminology regarding these objects on very Earth-like orbits with the hope of clarifying the literature and facilitating future research.

First at Mars and now at the Earth, the impact histories of the terrestrial planets appear to be partially encoded in the asteroids that orbit nearby to them. Future work to discover more of and measure the properties of this population of near-Earth objects which are sourced by the Moon will be critical to link asteroid and lunar science in the era of Artemis and the Vera Rubin Observatory's LSST.

\begin{acknowledgments}
We thank the staff of the Lowell Discovery Telescope and the NASA Infrared Telescope Facility for their help in obtaining these observations and the changes in schedule they required. In particular, we would like to thank Mike Connelly at the NASA IRTF for being critical to the success of those observations.

These results made use of the Lowell Discovery Telescope (LDT) at Lowell Observatory.  Lowell is a private, non-profit institution dedicated to astrophysical research and public appreciation of astronomy and operates the LDT in partnership with Boston University, the University of Maryland, the University of Toledo, Northern Arizona University and Yale University.

Visiting Astronomer at the Infrared Telescope Facility, which is operated by the University of Hawaii under contract 80HQTR24DA010 with the National Aeronautics and Space Administration. The authors wish to recognize and acknowledge the very significant cultural role and reverence that the summit of Maunakea has always had within the indigenous Hawaiian community. We are most fortunate to have the opportunity to conduct observations from this mountain.

N.M. and T.K. acknowledge funding from NASA YORPD grant 80NSSC21K1328 awarded in support of the Mission Accessible Near-Earth Object Survey (MANOS).

The work of O.F. was supported by an appointment to the NASA Postdoctoral Program at the Jet Propulsion Laboratory, administered by Oak Ridge Associated Universities under contract with NASA.

D.F. conducted this research at the Jet Propulsion Laboratory, California Institute of Technology, under a contract with the National Aeronautics and Space Administration (80NM0018D0004).

Taxonomic type results presented in this work were determined, in whole or in part, using a Bus-DeMeo Taxonomy Classification Web tool by Stephen M. Slivan, developed at MIT with the support of National Science Foundation Grant 0506716 and NASA Grant NAG5-12355.
\end{acknowledgments}

\vspace{5mm}
\facilities{LDT(LMI and DeVeny), IRTF(SpeX)}

\software{spextool, Pypeit}

\appendix
\section{Bus-Demeo Classification of lunar Spectra}
In this appendix, we expand on our analysis of laboratory spectra of lunar meteorites and returned samples. As described in Section \ref{sec:disc}, we obtained 863 samples of lunar materials with reflectance spectra covering $0.4-2.45$ $\mu{m}$ and then calculated the Root Mean Square (RMS) difference between each of the spectra and each of the classes in the Bus-Demeo asteroid taxonomy \citep{2009Icar..202..160D}. 45 of the 863 ($\approx5.2\%$) had at least one Bus-Demeo class to which their RMS difference was less than $0.2$, the closest `match' having $RMS=0.162$. As a consistency check, we took some of the best RMS matches and ingested them into the online Bus-Demeo classification tool \footnote{Avaialble at: http://smass.mit.edu/busdemeoclass.html}. That tool classifies spectra based on a Principle Component Analysis (PCA) but we retrieved nearly-identical taxonomic assignments for the variety of spectra tested. The few differences were in classifying spectra which were nearly equally matched by two different spectral classes (a difference in RMS of less than $0.01$) in which our preferred `best' assignment differed. As a broad overview, at relatively low SNR for telescopic spectra (perhaps $SNR = 10-20$), an object with a Moon-like reflectance spectrum could be identified as a relatively mundane asteroid maybe one time in twenty. At higher SNR, the chance of an object which reflects light like the lunar surface being identified as a standard asteroid begins to decrease as spectral features become better resolved.

For instance, some of the redder lunar spectra were equally well matched by the D and A type asteroids -- both quite red, but the D-types have no clear absorption features and the A-types have a very strong absorption around $1\mu{m}$. The lunar spectra fall intermediate between the two, with very red slopes and a weaker $1\mu{m}$ feature. The lunar spectra with blue overall slopes -- and thus the ones which were matched to the B-type asteroids -- also show weak $1\mu{m}$ features, which are are uncommon on carbonaceous asteroids generally. In other words, at moderate SNR or higher, a combined visible-and-near-infrared spectrum of lunar material will only rarely be matched to an asteroid class without some spectral feature making the match suspect. For the intermediate slope lunar spectra, the few that matched an asteroid class (12 of 863, $\approx1.4\%$ of the total) were assigned into the X and K types -- while there are some mismatches in their spectral behavior (in general, their continuum slopes are higher than typical objects in those classes but their band depths are larger, mimicking a lower spectral slope), these are harder to immediately identify as poor fits. If there are more lunar NEOs in the known-and-characterized population, these moderate-sloped X-class objects might be a fruitful area for future investigation. 

\begin{figure}[ht!]
\plotone{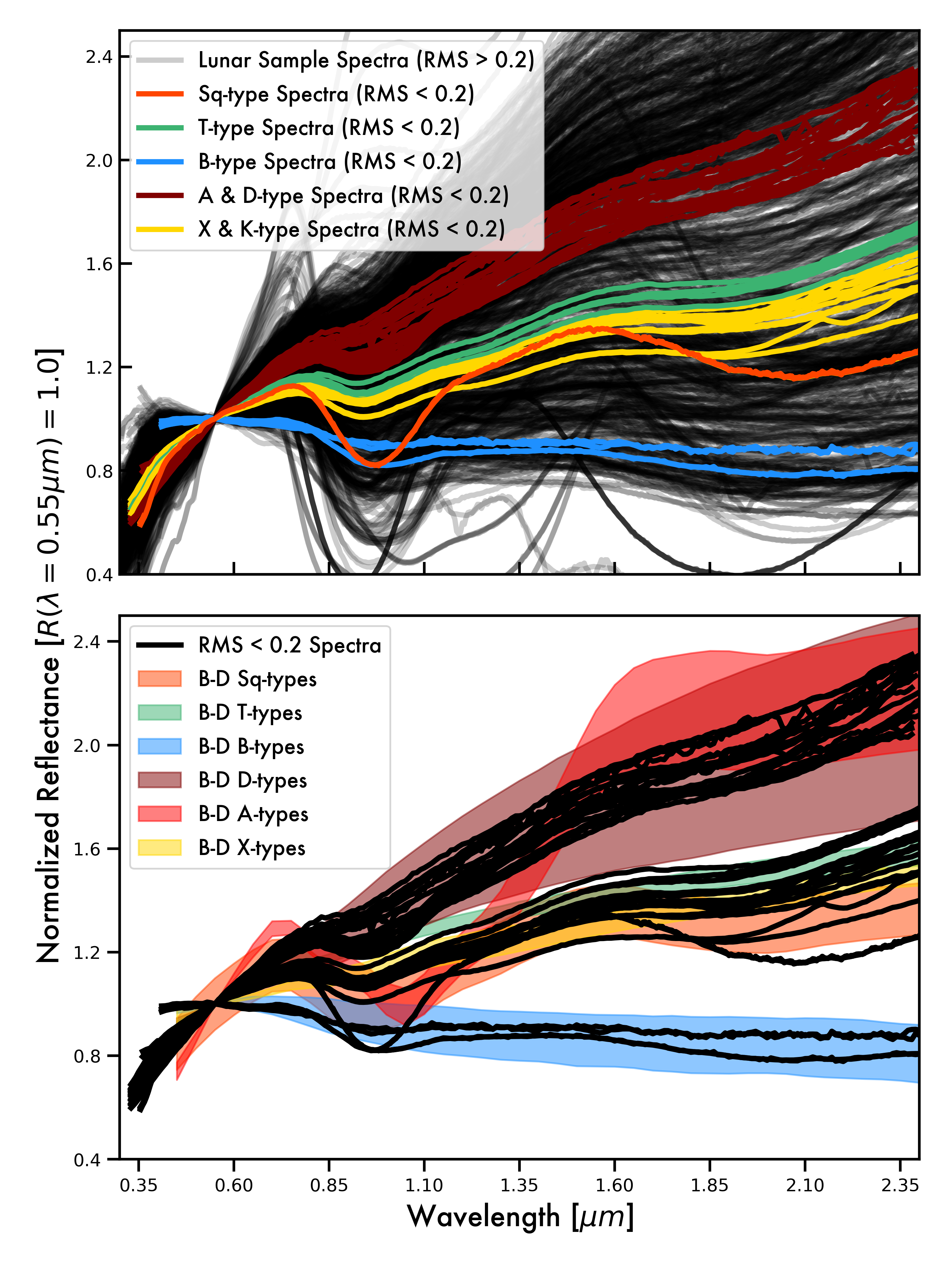}
\caption{Left: a comparison of 863 visible/near-infrared reflectance spectra of returned lunar samples and meteorites from RELAB. 45 of these spectra had an RMS less than $0.2$ when compared against a Bus-Demeo asteroid type, in which case they are colored according to which type they match the best. Right: The outlines of those same taxonomic classes are shown at right, with the `matched' spectra now plotted in black. With a few exceptions discussed further in the text, the vast majority of the `matched' spectra have features or slopes that would make them look like outliers within their respective taxa.} 
\label{fig:app_lunar_bus_demeo}
\end{figure}

As mentioned in the main text, analyzing visible or near-infrared reflectance spectra separately results in a higher fraction of lunar samples having close matches to the Bus-Demeo asteroid types. As can be seen from Figure \ref{fig:app_lunar_bus_demeo}, the range of visible slopes shown by lunar samples spans from redder than reddest asteroids to moderate blue slopes like many B-types. While some show absorption features close enough to visible wavelengths that a standard visible spectrum or set of colors (e.g., Sloan $g$, $r$, $i$, $z$) could identify it, some have their first absorption feature somewhat shallower and at longer wavelengths making even the identification of the surface as silicate rich and rocky challenging. In other words, attempting to identify if an object is lunar or asteroidal in origin based on just visible data alone is likely impossible. Near-infrared data on its own is less susceptible to these kinds of errors, but is still significantly less reliable than a visible/near-infrared approach. While the exact fraction of objects that might be classified as asteroidal varies considerably dependent on choice of wavelength range studied and what level of coarseness the spectra are binned to, only those with the absolute reddest slopes would look anomalous as asteroids based on visible data and those with the bluest and reddest slopes would look anomalous in in near-infrared data.

\section{Deciding on a lunar Origin}
This section summarizes the process that we recommend be followed to identify an object as \textit{potentially} lunar in origin as a guide for future observers and analysis. As stated in the main text, a combined process that utilizes both orbital and reflective information is almost certainly required. Furthermore, as discussed in the previous Section of this Appendix, information about how the object reflects light at both visible and near-infrared wavelengths is required to distinguish a lunar origin from asteroidal surfaces excepting the reddest lunar materials.

\begin{figure}[ht!]
\plotone{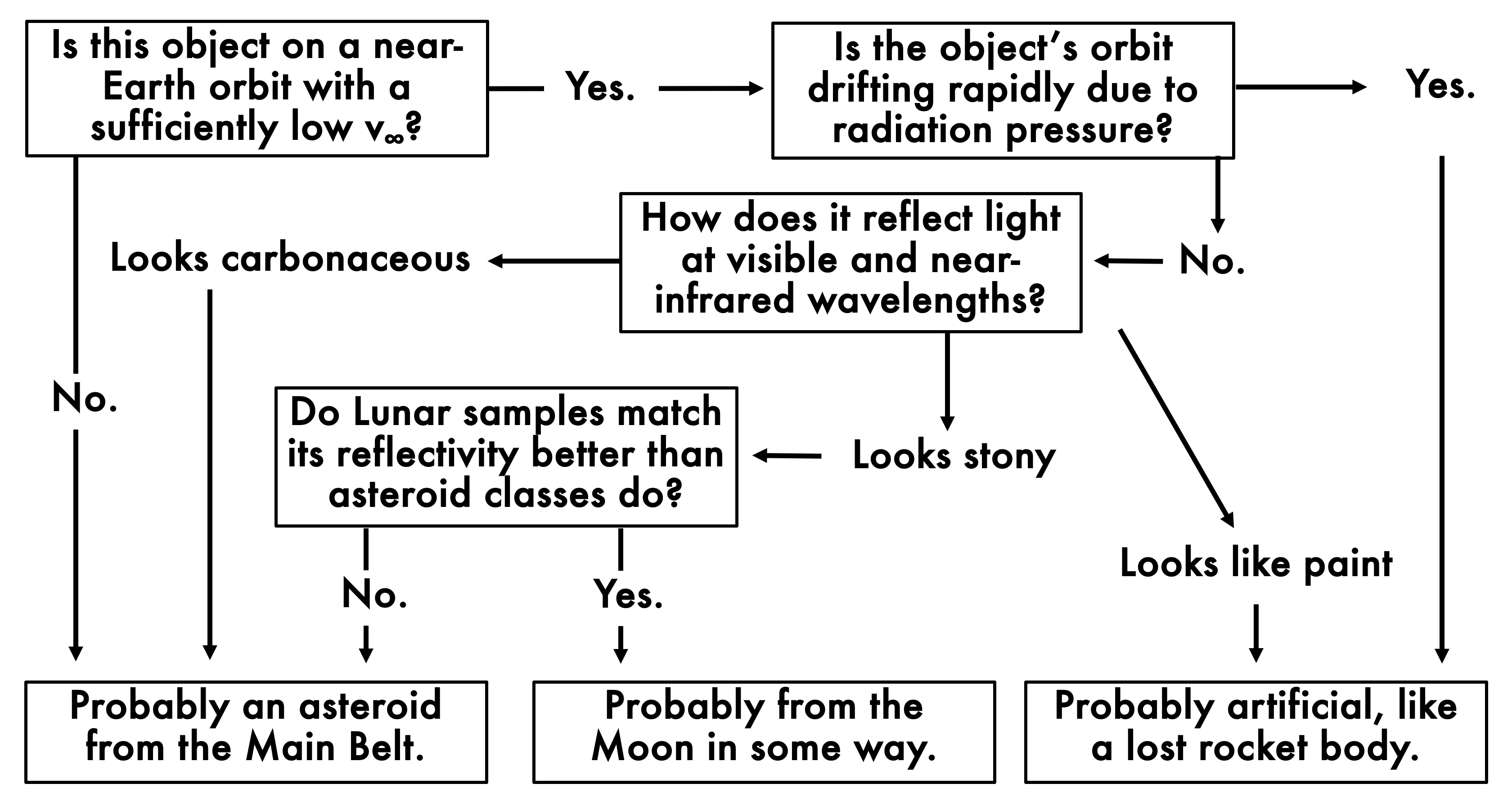}
\caption{A flowchart describing the general decision making process by which one might discern the origin of an object on a very Earth-like orbit based on its orbital and reflective properties. This approach will under-estimate the amount of lunar material in Near-Earth space, and the caveats to this simplified approach are discussed in the text.} 
\label{fig:app_lunar_flowchart}
\end{figure}

\begin{enumerate}
    \item Does this new near-Earth object have a low $v_\infty$? Lower values are better.
    \item If so, is the object's orbit significantly perturbed by radiation pressure -- and thus an indication that it might have a high Area-to-Mass ($A/m$) ratio?
    \item If not, does the reflectance spectrum of the object look like a solid natural surface at visible and near-infrared wavelengths?
    \item If so, does the surface look rocky with an absorption feature near $1.0-\mu{m}$? Band centers at shorter wavelengths ($<0.93\mu{m}$) are preferred based on the band centers of lunar spectra.
    \item If the surface looks natural and rocky, is it significantly easier to match the reflectivity of the object with returned samples and meteorites from the Moon as opposed to types from an asteroid taxonomic system like \citet{2009Icar..202..160D}?
\end{enumerate}

If the object is indeed low $v_\infty$, is not rapidly drifting due to radiation pressure, has a rocky surface, and generally reflects light significantly more like lunar samples than any rocky asteroid class, then it is likely to be derived from the Moon in some way. That said, there are several important caveats to be mentioned in this brief summary. First, material ejected from the Moon a long time ago may no longer be in such an Earth-like orbit -- this is likely the case for Kamo`oalewa, as it is in Quasi-Satellite like orbit around the Earth but does not formally have a low $v_\infty$. Some lunar material might be equally well fit by asteroidal classes and lunar samples, especially at low SNR -- and some asteroids will on occasion have ambiguous spectra as well. In other words, this general approach will \textit{under-estimate} the amount of lunar Ejecta in Near-Earth Space by discounting objects for which their origin is at least somewhat uncertain. For those kinds of objects, finding additional lines of evidence -- albedo from thermal measurements, more long-term or sophisticated dynamical studies -- would be critical to narrowing down the object's origins.


\begin{thebibliography}{}
\expandafter\ifx\csname natexlab\endcsname\relax\def\natexlab#1{#1}\fi
\providecommand{\url}[1]{\href{#1}{#1}}
\providecommand{\dodoi}[1]{doi:~\href{http://doi.org/#1}{\nolinkurl{#1}}}
\providecommand{\doeprint}[1]{\href{http://ascl.net/#1}{\nolinkurl{http://ascl.net/#1}}}
\providecommand{\doarXiv}[1]{\href{https://arxiv.org/abs/#1}{\nolinkurl{https://arxiv.org/abs/#1}}}

\bibitem[{{Battle} {et~al.}(2022){Battle}, {Reddy}, {Sanchez}, {Sharkey}, {Pearson}, \& {Bowen}}]{2022PSJ.....3..226B}
{Battle}, A., {Reddy}, V., {Sanchez}, J.~A., {et~al.} 2022, \psj, 3, 226, \dodoi{10.3847/PSJ/ac7223}

\bibitem[{{Battle} {et~al.}(2024){Battle}, {Reddy}, {Sanchez}, {Sharkey}, {Campbell}, {Chodas}, {Conrad}, {Engelhart}, {Frith}, {Furfaro}, {Farnocchia}, {Kuhn}, {Pearson}, {Rothberg}, {Veillet}, \& {Wainscoat}}]{2024PSJ.....5...96B}
---. 2024, \psj, 5, 96, \dodoi{10.3847/PSJ/ad3078}

\bibitem[{{Bida} {et~al.}(2014){Bida}, {Dunham}, {Massey}, \& {Roe}}]{2014SPIE.9147E..2NB}
{Bida}, T.~A., {Dunham}, E.~W., {Massey}, P., \& {Roe}, H.~G. 2014, in Society of Photo-Optical Instrumentation Engineers (SPIE) Conference Series, Vol. 9147, Ground-based and Airborne Instrumentation for Astronomy V, ed. S.~K. {Ramsay}, I.~S. {McLean}, \& H.~{Takami}, 91472N, \dodoi{10.1117/12.2056872}

\bibitem[{{Binzel} {et~al.}(2019){Binzel}, {DeMeo}, {Turtelboom}, {Bus}, {Tokunaga}, {Burbine}, {Lantz}, {Polishook}, {Carry}, {Morbidelli}, {Birlan}, {Vernazza}, {Burt}, {Moskovitz}, {Slivan}, {Thomas}, {Rivkin}, {Hicks}, {Dunn}, {Reddy}, {Sanchez}, {Granvik}, \& {Kohout}}]{2019Icar..324...41B}
{Binzel}, R.~P., {DeMeo}, F.~E., {Turtelboom}, E.~V., {et~al.} 2019, \icarus, 324, 41, \dodoi{10.1016/j.icarus.2018.12.035}

\bibitem[{{Bolin} {et~al.}(2024){Bolin}, {Denneau}, {Abron}, {Jedicke}, {Chiboucas}, {Ingerbretsen}, \& {Lemaux}}]{2024arXiv241108029B}
{Bolin}, B.~T., {Denneau}, L., {Abron}, L.-M., {et~al.} 2024, arXiv e-prints, arXiv:2411.08029.
\newblock \doarXiv{2411.08029}

\bibitem[{{Bottke} {et~al.}(2002){Bottke}, {Morbidelli}, {Jedicke}, {Petit}, {Levison}, {Michel}, \& {Metcalfe}}]{2002Icar..156..399B}
{Bottke}, W.~F., {Morbidelli}, A., {Jedicke}, R., {et~al.} 2002, \icarus, 156, 399, \dodoi{10.1006/icar.2001.6788}

\bibitem[{{Bowen} {et~al.}(2023){Bowen}, {Reddy}, {De Florio}, {Kareta}, {Pearson}, {Furfaro}, {Sharkey}, {McGraw}, {Cantillo}, {Sanchez}, \& {Battle}}]{2023PSJ.....4...52B}
{Bowen}, B., {Reddy}, V., {De Florio}, M., {et~al.} 2023, \psj, 4, 52, \dodoi{10.3847/PSJ/acb268}

\bibitem[{{Cantillo} {et~al.}(2023){Cantillo}, {Reddy}, {Battle}, {Sharkey}, {Pearson}, {Campbell}, {Satpathy}, {De Florio}, {Furfaro}, \& {Sanchez}}]{2023PSJ.....4..177C}
{Cantillo}, D.~C., {Reddy}, V., {Battle}, A., {et~al.} 2023, \psj, 4, 177, \dodoi{10.3847/PSJ/acf298}

\bibitem[{{Castro-Cisneros} {et~al.}(2023){Castro-Cisneros}, {Malhotra}, \& {Rosengren}}]{2023ComEE...4..372C}
{Castro-Cisneros}, J.~D., {Malhotra}, R., \& {Rosengren}, A.~J. 2023, Communications Earth and Environment, 4, 372, \dodoi{10.1038/s43247-023-01031-w}

\bibitem[{{Connors} {et~al.}(2004){Connors}, {Veillet}, {Brasser}, {Wiegert}, {Chodas}, {Mikkola}, \& {Innanen}}]{2004M&PS...39.1251C}
{Connors}, M., {Veillet}, C., {Brasser}, R., {et~al.} 2004, \maps, 39, 1251, \dodoi{10.1111/j.1945-5100.2004.tb00944.x}

\bibitem[{{Consolmagno} \& {Drake}(1977)}]{1977GeCoA..41.1271C}
{Consolmagno}, G.~J., \& {Drake}, M.~J. 1977, \gca, 41, 1271, \dodoi{10.1016/0016-7037(77)90072-2}

\bibitem[{{Cushing} {et~al.}(2004){Cushing}, {Vacca}, \& {Rayner}}]{2004PASP..116..362C}
{Cushing}, M.~C., {Vacca}, W.~D., \& {Rayner}, J.~T. 2004, \pasp, 116, 362, \dodoi{10.1086/382907}

\bibitem[{{de la Fuente Marcos} {et~al.}(2023){de la Fuente Marcos}, {de Le{\'o}n}, {de la Fuente Marcos}, {Licandro}, {Serra-Ricart}, \& {Cabrera-Lavers}}]{2023A&A...670L..10D}
{de la Fuente Marcos}, R., {de Le{\'o}n}, J., {de la Fuente Marcos}, C., {et~al.} 2023, \aap, 670, L10, \dodoi{10.1051/0004-6361/202245514}

\bibitem[{{DeMeo} {et~al.}(2009){DeMeo}, {Binzel}, {Slivan}, \& {Bus}}]{2009Icar..202..160D}
{DeMeo}, F.~E., {Binzel}, R.~P., {Slivan}, S.~M., \& {Bus}, S.~J. 2009, \icarus, 202, 160, \dodoi{10.1016/j.icarus.2009.02.005}

\bibitem[{{Devog{\`e}le} {et~al.}(2019){Devog{\`e}le}, {Moskovitz}, {Thirouin}, {Gustaffson}, {Magnuson}, {Thomas}, {Willman}, {Christensen}, {Person}, {Binzel}, {Polishook}, {DeMeo}, {Hinkle}, {Trilling}, {Mommert}, {Burt}, \& {Skiff}}]{2019AJ....158..196D}
{Devog{\`e}le}, M., {Moskovitz}, N., {Thirouin}, A., {et~al.} 2019, \aj, 158, 196, \dodoi{10.3847/1538-3881/ab43dd}

\bibitem[{{Farnocchia}(2021)}]{Farnocchia2021sb441}
{Farnocchia}, D. 2021, {Small-Body Perturber Files SB441-N16 and SB441-N343}, Tech. Rep. IOM 392R-21-005, Jet Propulsion Laboratory

\bibitem[{{Farnocchia} {et~al.}(2015){Farnocchia}, {Chesley}, {Milani}, {Gronchi}, \& {Chodas}}]{2015aste.book..815F}
{Farnocchia}, D., {Chesley}, S.~R., {Milani}, A., {Gronchi}, G.~F., \& {Chodas}, P.~W. 2015, in Asteroids IV (University of Arizona Press), 815--834, \dodoi{10.2458/azu_uapress_9780816532131-ch041}

\bibitem[{{Farnocchia} {et~al.}(2019){Farnocchia}, {Eggl}, {Chodas}, {Giorgini}, \& {Chesley}}]{2019CeMDA.131...36F}
{Farnocchia}, D., {Eggl}, S., {Chodas}, P.~W., {Giorgini}, J.~D., \& {Chesley}, S.~R. 2019, Celestial Mechanics and Dynamical Astronomy, 131, 36, \dodoi{10.1007/s10569-019-9914-4}

\bibitem[{{Giorgini} {et~al.}(2001){Giorgini}, {Chodas}, \& {Yeomans}}]{2001DPS....33.5813G}
{Giorgini}, J.~D., {Chodas}, P.~W., \& {Yeomans}, D.~K. 2001, in AAS/Division for Planetary Sciences Meeting Abstracts, Vol.~33, AAS/Division for Planetary Sciences Meeting Abstracts \#33, 58.13

\bibitem[{{Granvik} {et~al.}(2018){Granvik}, {Morbidelli}, {Jedicke}, {Bolin}, {Bottke}, {Beshore}, {Vokrouhlick{\'y}}, {Nesvorn{\'y}}, \& {Michel}}]{2018Icar..312..181G}
{Granvik}, M., {Morbidelli}, A., {Jedicke}, R., {et~al.} 2018, \icarus, 312, 181, \dodoi{10.1016/j.icarus.2018.04.018}

\bibitem[{{Higuchi} \& {Ida}(2017)}]{2017AJ....153..155H}
{Higuchi}, A., \& {Ida}, S. 2017, \aj, 153, 155, \dodoi{10.3847/1538-3881/aa5daa}

\bibitem[{{Jedicke} {et~al.}(2018){Jedicke}, {Bolin}, {Bottke}, {Chyba}, {Fedorets}, {Granvik}, {Jones}, \& {Urrutxua}}]{2018FrASS...5...13J}
{Jedicke}, R., {Bolin}, B.~T., {Bottke}, W.~F., {et~al.} 2018, Frontiers in Astronomy and Space Sciences, 5, 13, \dodoi{10.3389/fspas.2018.00013}

\bibitem[{Jenniskens {et~al.}(2016)Jenniskens, Albers, Koop, Odeh, Al-Noimy, Al-Remeithi, Hasmi, Dantowitz, Gasdia, Löhle, Zander, Hermann, Farnocchia, Chesley, Chodas, Park, Giorgini, Gray, Robertson, \& Lips}]{doi:10.2514/6.2016-0999}
Jenniskens, P., Albers, J., Koop, M.~W., {et~al.} 2016, Airborne observations of an asteroid entry for high fidelity modeling (AIAA), \dodoi{10.2514/6.2016-0999}

\bibitem[{{Jiao} {et~al.}(2024){Jiao}, {Cheng}, {Huang}, {Asphaug}, {Gladman}, {Malhotra}, {Michel}, {Yu}, \& {Baoyin}}]{2024NatAs...8..819J}
{Jiao}, Y., {Cheng}, B., {Huang}, Y., {et~al.} 2024, Nature Astronomy, 8, 819, \dodoi{10.1038/s41550-024-02258-z}

\bibitem[{{Kareta} {et~al.}(2022){Kareta}, {Reddy}, {Sanchez}, \& {Harris}}]{2022PSJ.....3..105K}
{Kareta}, T., {Reddy}, V., {Sanchez}, J.~A., \& {Harris}, W.~M. 2022, \psj, 3, 105, \dodoi{10.3847/PSJ/ac63cb}

\bibitem[{{Kareta} {et~al.}(2024){Kareta}, {Vida}, {Micheli}, {Moskovitz}, {Wiegert}, {Brown}, {McCausland}, {Devillepoix}, {Male{\v{c}}i{\'c}}, {Prtenjak}, {{\v{S}}egon}, {Shafransky}, \& {Farnocchia}}]{2024PSJ.....5..253K}
{Kareta}, T., {Vida}, D., {Micheli}, M., {et~al.} 2024, \psj, 5, 253, \dodoi{10.3847/PSJ/ad8b22}

\bibitem[{{Kary} \& {Dones}(1996)}]{1996Icar..121..207K}
{Kary}, D.~M., \& {Dones}, L. 1996, \icarus, 121, 207, \dodoi{10.1006/icar.1996.0082}

\bibitem[{{Kwiatkowski} {et~al.}(2009){Kwiatkowski}, {Kryszczy{\'n}ska}, {Poli{\'n}ska}, {Buckley}, {O'Donoghue}, {Charles}, {Crause}, {Crawford}, {Hashimoto}, {Kniazev}, {Loaring}, {Romero Colmenero}, {Sefako}, {Still}, \& {Vaisanen}}]{2009A&A...495..967K}
{Kwiatkowski}, T., {Kryszczy{\'n}ska}, A., {Poli{\'n}ska}, M., {et~al.} 2009, \aap, 495, 967, \dodoi{10.1051/0004-6361:200810965}

\bibitem[{Lin {et~al.}(2019)Lin, He, Yang, Lin, Xu, Zhang, Zhu, Chang, Zhang, Li, Lin, Liu, Gou, Wei, Hu, Xue, Yang, Zhong, Fu, Wan, \& Zou}]{10.1093/nsr/nwz183}
Lin, H., He, Z., Yang, W., {et~al.} 2019, National Science Review, 7, 913, \dodoi{10.1093/nsr/nwz183}

\bibitem[{{Luu} \& {Jewitt}(1990)}]{1990Icar...86...69L}
{Luu}, J., \& {Jewitt}, D. 1990, \icarus, 86, 69, \dodoi{10.1016/0019-1035(90)90199-J}

\bibitem[{{McCord} {et~al.}(1970){McCord}, {Adams}, \& {Johnson}}]{1970Sci...168.1445M}
{McCord}, T.~B., {Adams}, J.~B., \& {Johnson}, T.~V. 1970, Science, 168, 1445, \dodoi{10.1126/science.168.3938.1445}

\bibitem[{{Micheli} {et~al.}(2012){Micheli}, {Tholen}, \& {Elliott}}]{2012NewA...17..446M}
{Micheli}, M., {Tholen}, D.~J., \& {Elliott}, G.~T. 2012, \na, 17, 446, \dodoi{10.1016/j.newast.2011.11.008}

\bibitem[{{Micheli} {et~al.}(2013){Micheli}, {Tholen}, \& {Elliott}}]{2013Icar..226..251M}
---. 2013, \icarus, 226, 251, \dodoi{10.1016/j.icarus.2013.05.032}

\bibitem[{{Micheli} {et~al.}(2014){Micheli}, {Tholen}, \& {Elliott}}]{2014ApJ...788L...1M}
---. 2014, \apjl, 788, L1, \dodoi{10.1088/2041-8205/788/1/L1}

\bibitem[{{Mommert}(2017)}]{2017A&C....18...47M}
{Mommert}, M. 2017, Astronomy and Computing, 18, 47, \dodoi{10.1016/j.ascom.2016.11.002}

\bibitem[{{Naidu} {et~al.}(2021){Naidu}, {Micheli}, {Farnocchia}, {Roa}, {Fedorets}, {Christensen}, \& {Weryk}}]{2021ApJ...913L...6N}
{Naidu}, S.~P., {Micheli}, M., {Farnocchia}, D., {et~al.} 2021, \apjl, 913, L6, \dodoi{10.3847/2041-8213/abf836}

\bibitem[{{Nesvorn{\'y}} {et~al.}(2024){Nesvorn{\'y}}, {Vokrouhlick{\'y}}, {Shelly}, {Deienno}, {Bottke}, {Fuls}, {Jedicke}, {Naidu}, {Chesley}, {Chodas}, {Farnocchia}, \& {Delbo}}]{2024Icar..41716110N}
{Nesvorn{\'y}}, D., {Vokrouhlick{\'y}}, D., {Shelly}, F., {et~al.} 2024, \icarus, 417, 116110, \dodoi{10.1016/j.icarus.2024.116110}

\bibitem[{Park {et~al.}(2021)Park, Folkner, Williams, \& Boggs}]{Park2021}
Park, R.~S., Folkner, W.~M., Williams, J.~G., \& Boggs, D.~H. 2021, The Astronomical Journal, 161, \dodoi{10.3847/1538-3881/abd414}

\bibitem[{{Pieters}(1983)}]{1983JGR....88.9534P}
{Pieters}, C.~M. 1983, \jgr, 88, 9534, \dodoi{10.1029/JB088iB11p09534}

\bibitem[{{Polishook} {et~al.}(2017){Polishook}, {Jacobson}, {Morbidelli}, \& {Aharonson}}]{2017NatAs...1E.179P}
{Polishook}, D., {Jacobson}, S.~A., {Morbidelli}, A., \& {Aharonson}, O. 2017, Nature Astronomy, 1, 0179, \dodoi{10.1038/s41550-017-0179}

\bibitem[{{Prochaska} {et~al.}(2020){Prochaska}, {Hennawi}, {Westfall}, {Cooke}, {Wang}, {Hsyu}, {Davies}, {Farina}, \& {Pelliccia}}]{2020JOSS....5.2308P}
{Prochaska}, J., {Hennawi}, J., {Westfall}, K., {et~al.} 2020, The Journal of Open Source Software, 5, 2308, \dodoi{10.21105/joss.02308}

\bibitem[{{Qi} \& {Qiao}(2022)}]{2022AJ....163..211Q}
{Qi}, Y., \& {Qiao}, D. 2022, \aj, 163, 211, \dodoi{10.3847/1538-3881/ac5e2c}

\bibitem[{{Rayner} {et~al.}(2003){Rayner}, {Toomey}, {Onaka}, {Denault}, {Stahlberger}, {Vacca}, {Cushing}, \& {Wang}}]{2003PASP..115..362R}
{Rayner}, J.~T., {Toomey}, D.~W., {Onaka}, P.~M., {et~al.} 2003, \pasp, 115, 362, \dodoi{10.1086/367745}

\bibitem[{{Reddy} {et~al.}(2009){Reddy}, {Emery}, {Gaffey}, {Bottke}, {Cramer}, \& {Kelley}}]{2009M&PS...44.1917R}
{Reddy}, V., {Emery}, J.~P., {Gaffey}, M.~J., {et~al.} 2009, \maps, 44, 1917, \dodoi{10.1111/j.1945-5100.2009.tb02001.x}

\bibitem[{{Rivkin} {et~al.}(2005){Rivkin}, {Binzel}, \& {Bus}}]{2005Icar..175..175R}
{Rivkin}, A.~S., {Binzel}, R.~P., \& {Bus}, S.~J. 2005, \icarus, 175, 175, \dodoi{10.1016/j.icarus.2004.11.005}

\bibitem[{{Sanchez} {et~al.}(2012){Sanchez}, {Reddy}, {Nathues}, {Cloutis}, {Mann}, \& {Hiesinger}}]{2012Icar..220...36S}
{Sanchez}, J.~A., {Reddy}, V., {Nathues}, A., {et~al.} 2012, \icarus, 220, 36, \dodoi{10.1016/j.icarus.2012.04.008}

\bibitem[{{Sanchez} {et~al.}(2020){Sanchez}, {Thomas}, {Reddy}, {Frere}, {Lindsay}, \& {Mitchell}}]{2020AJ....159..146S}
{Sanchez}, J.~A., {Thomas}, C., {Reddy}, V., {et~al.} 2020, \aj, 159, 146, \dodoi{10.3847/1538-3881/ab723f}

\bibitem[{{Sanchez} {et~al.}(2024){Sanchez}, {Reddy}, {Thirouin}, {Bottke}, {Kareta}, {De Florio}, {Sharkey}, {Battle}, {Cantillo}, \& {Pearson}}]{2024PSJ.....5..131S}
{Sanchez}, J.~A., {Reddy}, V., {Thirouin}, A., {et~al.} 2024, \psj, 5, 131, \dodoi{10.3847/PSJ/ad445f}

\bibitem[{{Sharkey} {et~al.}(2021){Sharkey}, {Reddy}, {Malhotra}, {Thirouin}, {Kuhn}, {Conrad}, {Rothberg}, {Sanchez}, {Thompson}, \& {Veillet}}]{2021ComEE...2..231S}
{Sharkey}, B. N.~L., {Reddy}, V., {Malhotra}, R., {et~al.} 2021, Communications Earth and Environment, 2, 231, \dodoi{10.1038/s43247-021-00303-7}

\bibitem[{{Speyerer} {et~al.}(2016){Speyerer}, {Povilaitis}, {Robinson}, {Thomas}, \& {Wagner}}]{2016Natur.538..215S}
{Speyerer}, E.~J., {Povilaitis}, R.~Z., {Robinson}, M.~S., {Thomas}, P.~C., \& {Wagner}, R.~V. 2016, \nat, 538, 215, \dodoi{10.1038/nature19829}

\bibitem[{{Thirouin} {et~al.}(2016){Thirouin}, {Moskovitz}, {Binzel}, {Christensen}, {DeMeo}, {Person}, {Polishook}, {Thomas}, {Trilling}, {Willman}, {Hinkle}, {Burt}, {Avner}, \& {Aceituno}}]{2016AJ....152..163T}
{Thirouin}, A., {Moskovitz}, N., {Binzel}, R.~P., {et~al.} 2016, \aj, 152, 163, \dodoi{10.3847/0004-6256/152/6/163}

\bibitem[{{Tonry} {et~al.}(2024){Tonry}, {Robinson}, {Fitzsimmons}, {Denneau}, {Weiland}, {Siverd}, {Erasmus}, {Tonry}, {Robinson}, {Fitzsimmons}, {Denneau}, {Weiland}, {Siverd}, {Erasmus}, {Tonry}, {Robinson}, {Fitzsimmons}, {Denneau}, {Weiland}, {Siverd}, {Erasmus}, {Losse}, {Ursache}, {Birtwhistle}, {Hudin}, {Sannino}, {Agnetti}, {Mercanti}, {Mancuso}, {Wells}, {Galli}, {Buzzi}, {Pimentel}, {Caetano}, {Jacques}, {Parrott}, {Rogers}, {Haeusler}, {Tonry}, {Robinson}, {Fitzsimmons}, {Denneau}, {Weiland}, {Siverd}, {Erasmus}, {Watanabe}, {Korlevic}, {Pettarin}, {Parrott}, {Kuettner}, {Haeusler}, {Maestripieri}, {Fagioli}, {Bacci}, {Tesi}, {Lombardo}, {Bernardi}, {Cheli}, {Squilloni}, {Grazzini}, {Mazzanti}, {Tombelli}, {Bitossi}, {Ceccarini}, {Caprara}, {Venticinque}, {Forti Parri}, {Gambacciani}, {Iozzi}, \& {Urbanik}}]{2024MPEC....P..170T}
{Tonry}, J., {Robinson}, J., {Fitzsimmons}, A., {et~al.} 2024, Minor Planet Electronic Circulars, 2024-P170, \dodoi{10.48377/MPEC/2024-P170}

\bibitem[{{Tonry} {et~al.}(2018){Tonry}, {Denneau}, {Heinze}, {Stalder}, {Smith}, {Smartt}, {Stubbs}, {Weiland}, \& {Rest}}]{2018PASP..130f4505T}
{Tonry}, J.~L., {Denneau}, L., {Heinze}, A.~N., {et~al.} 2018, \pasp, 130, 064505, \dodoi{10.1088/1538-3873/aabadf}

\bibitem[{{Valsecchi} {et~al.}(2015){Valsecchi}, {Alessi}, \& {Rossi}}]{2015CeMDA.123..151V}
{Valsecchi}, G.~B., {Alessi}, E.~M., \& {Rossi}, A. 2015, Celestial Mechanics and Dynamical Astronomy, 123, 151, \dodoi{10.1007/s10569-015-9631-6}

\bibitem[{{Valsecchi} {et~al.}(2003){Valsecchi}, {Milani}, {Gronchi}, \& {Chesley}}]{2003A&A...408.1179V}
{Valsecchi}, G.~B., {Milani}, A., {Gronchi}, G.~F., \& {Chesley}, S.~R. 2003, \aap, 408, 1179, \dodoi{10.1051/0004-6361:20031039}

\bibitem[{{Willmer}(2018)}]{2018ApJS..236...47W}
{Willmer}, C. N.~A. 2018, \apjs, 236, 47, \dodoi{10.3847/1538-4365/aabfdf}

\end{thebibliography}

\bibliographystyle{aasjournal}

\end{document}